\pdfoutput=1
\documentclass[preprint2]{aastex}
\usepackage{natbib}
\usepackage{amssymb}
\usepackage{multirow}
\usepackage{rotating}
\usepackage{longtable}
\usepackage{lscape}
\shorttitle{HI in Galaxy Groups}
\shortauthors{Chynoweth et al.}

\begin{document}

\title{The Origin of Neutral Hydrogen Clouds in Nearby Galaxy Groups: Exploring the Range Of Galaxy Interactions}

\author{Katie M. Chynoweth}  
\affil{National Research Council Postdoctoral Research Associate}
\affil{Resident at Naval Research Laboratory,
Washington, DC 20375}
\author{Kelly Holley-Bockelmann}  
\affil{Vanderbilt University, Physics and Astronomy Department, 1807 Station B, 
Nashville, TN 37235}
\affil{Fisk University, Department of Physics, Nashville, TN 37208}
\author{Emil Polisensky}
\affil{Naval Research Laboratory,
Washington, DC 20375}
\author{Glen I. Langston}
\affil{National Radio Astronomy Observatory,
Green Bank, WV 24944}

\begin{abstract}

We combine high resolution N-body simulations with deep observations of neutral hydrogen (HI) in nearby galaxy 
groups in order to explore two well-known theories of HI cloud formation: HI stripping by galaxy interactions and dark matter minihalos with embedded HI gas. 
This paper presents new data from three galaxy groups, Canes Venatici I, NGC 672, and NGC 45, and assembles data from our previous galaxy group campaign to generate a rich HI cloud archive to compare to our simulated data. 

We find no HI clouds in the Canes Venatici I, NGC 672, or NGC 45 galaxy groups. We conclude that HI clouds in our detection space are most likely to be generated through recent, strong galaxy interactions. We find no evidence of HI clouds associated with dark matter halos above M$_{HI}$ $\sim$ 10$^6$ M$_{\odot}$, within $\pm$ 700 km s$^{-1}$ of galaxies, and within 50 kpc projected distance of galaxies.

\end{abstract}


\section{Introduction}

Neutral hydrogen observations of the Milky Way  reveal a
repository of cold gas clouds that do not appear to rotate with the
galaxy. Since they are kinematically decoupled from galactic rotation,
they are known as High Velocity Clouds (HVCs). Generally, a cloud
is categorized as a HVC if its velocity deviates by more than 50 km s$^{-1}$ from the range allowed by a simple model of differential galactic rotation \citep{wvw91}.

Milky Way HVCs were first discovered by \citet{muller}, after detection of interstellar clouds through absorption line studies prompted systematic searches for HI clouds in equilibrium with a Galactic corona \citep{wvw97}. Subsequent studies have revealed that the demographics of HVCs in the Milky Way are rich and varied: HVCs are both ionized and neutral, occur in long streams and spheroidal clouds, and are found in large complexes and as small isolated clouds. The neutral hydrogen HVCs cover nearly 40\% of the sky \citep{wvw97, loc02}. In contrast, the ionized clouds cover as much as 85\% of the sky \citep{shull09}. In total, these clouds make up as much as 10\% of the total HI mass of the Milky Way \citep{wvw97}.

Possible analogs to the Milky Way clouds have been found around nearby galaxies including M31 \citep{thilker04}, M33 \citep{grossi08}, and others \citep{vdH88, miller09, hess09}. The best-studied external HI cloud system in an isolated galaxy is that of M31.
Around the galaxy, \citet{west08} found a population of HI clouds with properties consistent with those of Milky Way HVCs. The M31 clouds have typical HI masses of 10$^5$ M$_{\odot}$, and the system extends 50 kpc from the M31 disk.

There are several proposed theories for the source of HI clouds. Broadly, they
consist of two classes: those that invoke a galactic plane origin and rely on secular dynamics to remove the gas from the disk, and those that invoke an extraplanar origin and tie the dynamics to large scale structure.  Table \ref{millertable} lists the expected properties of HI clouds in each case, which we describe below. Note that the properties of cold-accretion generated clouds are not well-constrained by current simulations, particularly in terms of their velocities and sizes. 

In a galactic fountain, gas is blown out of the disk by a supernova
outburst, after which it cools, condenses and rains back onto the disk \citep{shap76, bre80}.  The galactic fountain model for HVCs is supported by the existence of a hot  (10$^{6}$ K) galactic corona with a scale height of a few kpc, detected in x-rays around galaxies (e.g., \citet{strick04}). This corona is supplied with gas by supernovae explosions from the disk, and radiative cooling causes the formation of HI clouds through thermal instabilities \citep{shap76, bre80}.  Although galactic fountains most certainly exist at some level in all star-forming galaxies, not all observed HI clouds are produced this way \citep{frat06}. In particular, the galactic fountain model fails to explain HI clouds further than a few kpc from the disk of a galaxy, beyond the extent of the hot galactic corona. 
Our observations are not sensitive to galactic fountain-generated HI clouds, as we are limited by both angular resolution and mass sensitivity.

Galaxy interactions may generate HI clouds through tidal stripping ~\citep{put03, chy08, san08}.  In the Milky Way, the Magellanic Stream is a prime example of this phenomenon--here gas was stripped out of a satellite galaxy during a minor merger. Another possible source of HI clouds far from galaxy disks is leftover fragments of tidal tails.

HI clouds may be tracers of a population of
~$10^7-10^{10} M_\odot$ dark matter halos~\citep{kra04}. Cosmological simulations based upon the accepted paradigm of $\Lambda$CDM (Cold Dark Matter) cosmology, such as the Via Lactea II simulation of \citet{die08}, consistently predict a high degree of dark matter substructure.  This aspect of simulations is not well-matched to observations--for instance, the Milky Way is observed to have a factor of at least 4 fewer satellites (in the form of dwarf galaxies) than predicted by simulations, despite sophisticated treatment of baryonic heating, cooling, and feedback in low mass halos \citep{sim07}. One possible solution to this ``missing satellites problem" could be HI clouds, if such objects are tracers of dark matter halos. However, there are still not enough clouds observed to supply the predicted number of satellites, and the clouds that are observed do not have the expected spatial or kinematic distribution of dark matter halos \citep{chy09,chy11}.

Numerical simulations show that gas may arrive to galaxies via accretion along dark matter filaments \citep{keres05}. Gas accretion along filaments appears to happen in two modes, a ``hot" (T $\sim$ 10$^6$ K) and a ``cold"  (T $<$ 10$^5$ K) mode.  The cold mode of accretion, where gas is not shock heated to the halo virial temperature as it falls in, is of primary relevance to HI studies, since some cold-mode accretion gas may appear as HI \citep{keres09}. HI clouds could be a result of instabilities in a hot halo that causes the gas to cool in clumps and fall
toward the galactic midplane~\citep{mal04,keres09}.  In this case,
HI clouds are cold, pressure-confined clouds without an associated dark matter
halo~\citep{keres09}. When cold gas encounters the ambient hot halo gas in a galaxy, a density inversion develops which causes a Rayleigh-Taylor instability.  This scenario is supported by the kinematics of the Milky Way HVC system, which imply significant inflow \citep{mir84,simon06,wakker07}.

Determining the relative importance of each HI cloud formation channel requires us to know the phase space and mass distribution of the clouds, and
measurement of the physical properties of HI clouds is dependent on an accurate value for distance.  For example, mass scales as D$^2$, size as D$^1$, and density and pressure as D$^{-1}$, where D represents distance \citep{wvw97}. Unfortunately, the distances to the clouds in the Milky Way are very difficult to measure. HVCs are starless, so none of the traditional methods for measuring the distance to a star can be used. The clouds' anomalous velocities and extraplanar positions make it impossible to use a model of disk rotation to determine their distances. The best that can be done in the Milky Way is to use stellar absorption lines to determine upper and lower distance limits for a cloud \citep{sch95, wvw97}.  In order to determine the upper distance limit, absorption in the spectrum of a star at the velocity of the cloud must be detected in a star in the line-of-sight of the cloud. For a lower distance limit, a significant non-detection of the absorption line in a star in the line-of-sight is required.  Many attempts have been made to measure the distances of clouds in the Milky Way using stellar absorption (see, e.g., \citet{sch95,vanw99, bekhti06}). These studies have placed cloud distances in the range of 5-10 kpc above and below the disk \citep{bar08}. Because background stars are required for this method of distance measurement, only clouds that are relatively close to the disk can be measured.

 An advantage to studying HI clouds in external galaxies is that there are many robust methods for measuring the distance to a galaxy, so the distance to any associated HI cloud is known much more accurately. Additionally, the projected distance between a cloud and galaxy is easy to measure for extragalactic HI clouds.
 Galaxies are most commonly found in groups (e.g. \citet{gel83,tul87, eke04, tago08}), so groups are logical environments in which to search for HI clouds. One HI cloud search of galaxy groups has been done -- \citet{pis07} searched a sample of loose groups that are Local Group analogs. They did not find any HI clouds, but used their observations to set the constraint that HI clouds should be on average clustered within 90 kpc of group galaxies, with an average mass of 4 $\times$ 10$^5$ M$_{\odot}$.

In \citet{chy08}, we presented observations of the M81 group. In \citet{chy09}, we investigated the NGC 2403 group. In \citet{chy11},  we expanded the observations of \citet{chy08} and \citet{chy09} to include the the filament between M81 and NGC 2403. In this paper, we search three more groups and synthesize this data in the context of our previous results.

This paper is organized as follows. In Section \ref{sample}, we describe our sample of galaxy groups and method of selection,  introduce our method of quantifying galaxy interaction strength. In Section \ref{obsec}, we describe the observation strategy and data reduction process. Section \ref{complete} presents a discussion on the completeness of the study. Section \ref{simulations} describes the predictions of numerical simulations for HI clouds associated with dark matter halos. Finally, in Section \ref{consec} we present our results and conclusions, and discuss future work.

\section{Sample Selection}
\label{sample}

We observed five
nearby galaxy groups with the 100m Robert C. Byrd Green Bank Telescope (GBT) at the NRAO\footnote{The National Radio Astronomy Observatory (NRAO) is a facility of the National Science Foundation operated under cooperative agreement by Associated Universities, Inc.} in Green Bank, West Virginia. We observed a wide angular area and velocity range around each galaxy, and the properties of detected HI clouds were catalogued and compared against simulations.

The five galaxy groups were chosen with the following criteria.
\begin{enumerate}
\item{Galaxy groups must span the range in intensity of galaxy interactions.}
\item{Galaxy groups must be nearby, for adequate angular resolution and sensitivity. We aimed for a maximum group distance of 7.5 Mpc, corresponding to an angular resolution of 20 kpc.}
\item{Galaxy groups must be at least partially distinct from the Milky Way in velocity space to avoid confusion with foreground HI gas. }
\end{enumerate}

Criterion 1 is essential in order to explore the role of interactions in the generation of HI clouds. We attempted to choose galaxy groups with different levels of galaxy interactions.  Although in some cases it is obvious that a galaxy interaction is taking place, it is not always a simple matter to determine whether a particular galaxy has undergone a recent interaction \citep{cons08,dar10}. Determining the relative strength or intensity of interactions is even more difficult.  However, we have made a first attempt towards quantifying galaxy interaction strength within galaxy groups by creating ``interaction indices", which we have applied to our galaxy groups in order to rank them in order of interaction strength (see Section \ref{interactions}).

\begin{figure}[h!]
\centering  
\includegraphics[width=5in]{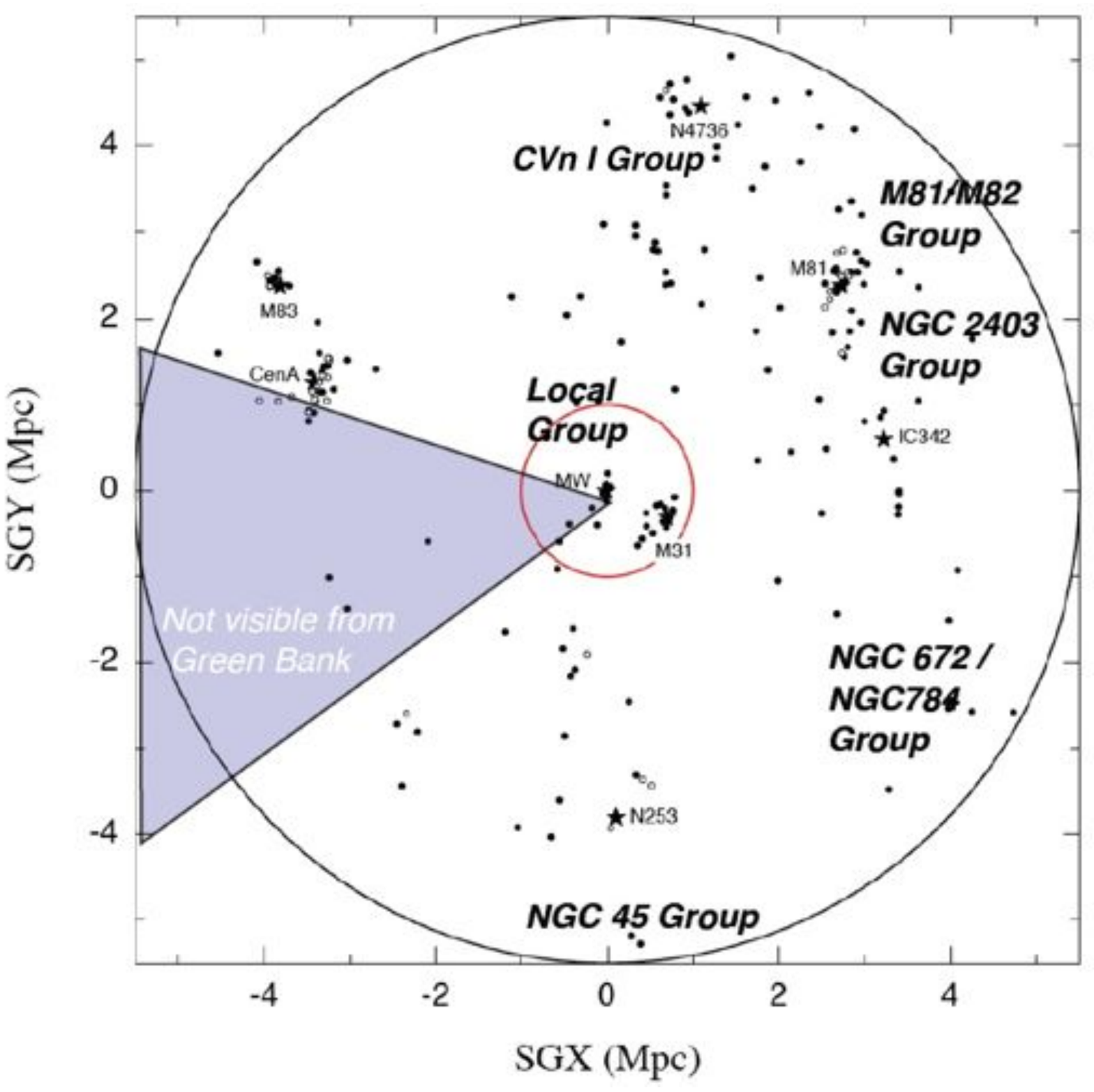}
\caption[Map of galaxy group locations]{Map of observed galaxy group locations. The gray wedge indicates the declination range not observable from Green Bank.} \label{fig:groups}
\end{figure}

Criteria 2 and 3 are difficult to reconcile, so the two closest groups (M81 and NGC 2403) partially overlap with the Milky Way in velocity space. The groups discussed in this paper do not overlap in velocity with the Milky Way.  The galaxy groups observed for this study are described below. Table \ref{tenmpc} lists all galaxy groups within 7.5 Mpc \citep{fou92,tul87}. Our observations cover 63\% of galaxy groups within 7.5 Mpc. Figure \ref{fig:groups} shows a diagram of the distribution of the observed groups.  Here we list the groups.

\subsection{M81 Group}

The M81 group is one of the closest galaxy groups to the Milky Way at a distance
of 3.5 Mpc. It is comprised of 4 major galaxies (M81, M82, NGC 3077, NGC 2976) and over 40 dwarf galaxies. Although the galaxies are optically undisturbed, HI observations reveal that the galaxies are undergoing strong interactions. Additionally, M82 is an extreme starburst galaxy and many of the dwarf galaxies show morphological and kinematic irregularities. Table \ref{m81tab} lists galaxy properties.

\subsection{NGC 2403 Group}

The NGC 2403 group is located at approximately the same distance as the M81
group. The NGC 2403 group contains the bright spiral galaxy NGC 2403, 3 additional large galaxies (NGC 2366, UGC 4305, UGC 4483), and at least 4 dwarf or low surface brightness satellite galaxies. The NGC 2403 group does not show clear signs of interaction. However, some of the group galaxies show subtle irregularities which may be due to previous interactions between group galaxies. The largest galaxy, NGC 2403, shows a thick, lagging HI layer and a few HI filaments with anomalous velocities that are similar to the M81/M82 clouds in mass \citep{frat02}. These features may be analogues to Milky Way HVCs, and their origins are similarly unclear. They have been attributed to a combination of outflow from a galactic fountain and accretion from the intergalactic medium \citep{frat06}; both of these phenomena may be induced by interactions between group galaxies \citep{lar78,san08}. UGC 4305 has a cometary appearance in HI \citep{bureau02}, which may be caused by ram pressure from the intragroup medium. The dwarf galaxy DDO 53 is also kinematically irregular in HI \citep{beg06}. The other galaxies in the group appear to be normal. Table \ref{n2403tab} lists galaxy properties.

\subsection{Canes Venatici I Group}

The Canes Venatici I (CVn I) galaxy group is located approximately 4 Mpc away and contains 41
galaxies distributed in a diffuse cloud. The brightest galaxy, M94,
is apparently isolated but is morphologically disturbed and has a ring \citep{mul93}. Five of the galaxies show bursts of star formation \citep{kk08}. Table \ref{canes1tab} lists galaxy properties.

\subsection{NGC 672 Group}

The NGC 672 and NGC 784 groups form a 6$^{\circ}$ linear filament of galaxies at a
distance of approximately 5 Mpc \citep{zit08}. The galaxies in both groups show evidence that a burst of star formation occurred approximately 10 Myr ago. \citet{zit08} conclude that these coincident SF bursts are most likely not due to galaxy interactions, since the SFR of the galaxies does not correlate with projected distance from the brightest and most massive galaxy. Instead, they attribute the bursts in star formation to accretion along the filament. However, it is well known that NGC 672 and IC 1727 are undergoing a strong interaction \citep{com80, soh96}. Table \ref{n672tab} lists galaxy properties.

\subsection{NGC 45 Group}

The NGC 45 group contains 3 large, low surface brightness galaxies (NGC 45,
NGC 24, NGC 59) and is located at a distance of approximately 6 Mpc \citep{fou92}. NGC 45 has a kinematical twist in the major axis, but NGC 24 is kinematically regular \citep{che06}. However, NGC 24 has star forming regions scattered across the entire disk \citep{ros03}. NGC 59 is the third galaxy in the group, and it appears to be normal and undisturbed. Table \ref{n45tab} lists galaxy properties.

\subsection{Quantifying Galaxy Interactions}
\label{interactions}

Measuring the probability of a recent galaxy interaction or merger in an observational dataset has been a goal for decades; common methods employ pair counts \citep{zep89, carl94, pat97, lef00, pat02, con03, bell06, berrier06}, morphological classification techniques (i.e., concentration, asymmetry, and clumpiness \citep{abraham96,con03} or the Gini coefficient \citep{lotz04}), or star formation \citep{sand88,kna09}. None of these are iron-clad merger indicators-- phenomena that are sometimes induced by interactions are not a guarantee of interactions. Furthermore, it is not clear how much weight to give each possible indicator. With these caveats, we used a combination of these parameters as a guide to rank the level of interactions in a galaxy group. 
Here we describe how we compiled merger characteristics into an ``interaction index". We use this as a rough guide to quantifying galaxy interactions.

A key symptom of galaxy interaction is found in the galaxy morphology. 
Various morphological features may be induced by galaxy interactions.  Although spiral, barred, and elliptical galaxies can all be related to interactions, the most reliable morphological indication of recent interactions are tidal bridges, tails, and shells. 

Secondly, strong central star formation is a key feature of recent or ongoing interactions.
Many interacting systems show indications of starburst activity such as high levels of H$\alpha$ emission and high infrared luminosity (e.g., \citet{sand88,kna09}). However, not all interacting systems show enhanced star-forming activity. The effect of orbital parameters and internal structure of the galaxies play an important part in the details of gas dynamics leading to star formation (see, e.g., \citet{cox08,mh96}). The detailed relationship between galaxy interactions and star formation is not well understood. However, in general it is found that the star formation rate (SFR) is increased in merging spiral galaxies (e.g., \citet{kenn84, keel85, bus86, hum90, darg10}). Therefore, we use the relative star formation rates in our observed sample as an indicator of ongoing galaxy interactions.

Finally, AGN activity may be triggered by galaxy interactions. This is because a huge amount of gas must be funneled to the nuclear regions to fuel the AGN's activity. There are various physical mechanisms by which gas loses angular momentum and falls to the center of a galaxy, including gravitational torques, viscous torques, and hydrodynamical torques or shocks. Galaxy interactions and mergers provide efficient mechanisms for funneling gas to the nucleus of a galaxy \citep{mh96}.
There are links between galaxy interactions and the onset of nuclear activity in galaxies in both observations \citep{sand88} and simulations \citep{spri05}. Many active galaxies show signs of interaction and merger such as disturbed morphology, particularly in HI \citep{kuo08}. Seyfert nuclei are found preferentially in interacting galaxies \citep{kenn84, lau95}. The strongest links between interactions and AGN are found in the systems with the highest luminosity \citep{bah97}.

Using these three factors, we calculate an Òinteraction indexÓ, described below. The index includes only those galaxies within $\pm$ 1 Mpc of the most massive galaxy in each group. For galaxies with no distance measurement, the distance of the most massive galaxy in the group was assumed. The index is applied to the sample of galaxy groups studied for this paper as well as the M81 and NGC 2403 groups and the Local Group. Note that the numbers resulting from these calculations are arbitrary; in order to use the index to rank our observation sample, we have normalized each variable to the maximum for our sample. A robust statistic would need to include a normalization of the indices to some standard and appropriate weighting of each variable. Fortunately, for the purposes of this study it is only necessary to rank the groups relative to one another.

We define the evidence for recent or ongoing interactions as follows:

\begin{equation}
\rm{I} =	\frac{N_{Tidal}}{max(N_{Tidal})} + \frac{SFR}{max(SFR)}	+ \frac{N_{AGN}}{max(N_{AGN})},
\end{equation}

\noindent Here, the first factor describes the degree of disturbed morphology; the second factor the star formation rates, and the third factor accounts for AGN activity. $N_{Tidal}$ is the number of group galaxies exhibiting tidal tails or bridges. $SFR$ is the average star formation rate of group galaxies, calculated following \citet{ken98}, using Galactic extinction-corrected H$\alpha$ luminosities 11HUGS study of \citet{ken08}.  For galaxies with no H$\alpha$ luminosity measurement in the 11HUGS survey \citep{ken08}, the 11HUGS survey detection limit was assumed. $N_{\rm{AGN}}$ is the number of group galaxies classified as any type of AGN in NED\footnote{NASA/IPAC Extragalactic Database (NED) which is operated by the Jet Propulsion Laboratory, California Institute of Technology, under contract with the National Aeronautics and Space Administration, http://nedwww.ipac.caltech.edu/}.

We applied Equation (1) to the five galaxy groups. The results are tabulated in Tables \ref{ind_ev_val} and \ref{ind_ev_rank}.  To avoid giving any one factor an artificially high or low weight, we assign a rank from 1-6 for each variable, and add the ranks for the final ranking order for our sample.  Since this ranking scheme is, at best, an ad-hoc indicator of ongoing interactions, we also tested alternate versions of Equation (1). We tested each permutation of two of the three factors, as well as each factor individually, and each version that we tested yielded a slightly different result. Since none of the factors is an iron-clad indicator of ongoing interactions in and of itself, we concluded that it is best to use all available evidence in ranking the galaxy groups with respect to one another.

The M81 group has the highest interaction index, while the NGC 45 group ranks lowest.  Based upon these rankings, we expect that the M81 group should contain the most HI clouds, followed by NGC 672, CVn I, NGC 2403, and NGC 45.
To place these calculations in context, we have also calculated the interaction index for the Local Group. The Local Group measurements are included in boldface in Tables \ref{ind_ev_val} and \ref{ind_ev_rank}. We consider the Local Group to consist of the Milky Way, M33, and M31 since Local Group dwarf galaxies would not be visible from the distances of the groups we observed. We calculate the M31 and M33 star formation rate from the 11HUGS survey H$\alpha$ luminosity, and assume the widely-accepted star formation rate of 1 M$_{\odot}$ yr$^{-1}$ for the Milky Way. The Local Group ranks second to the M81 group in interaction evidence.

We wish to point out a few notable areas where our ranking scheme is lacking. Firstly, we have not attempted to put a quantitative timescale on the index. This is because the factors we take into account have different (and largely unknown) durations, especially in relation to interactions, and combining these different timescales is a daunting problem. Secondly, we have no method in place for predicting the quantity of HI clouds that ought to be present or their properties based on the interaction indices. Rather, we predict whether a given galaxy group is more likely than another to host HI clouds at all. Finally, we have not considered possible cross-correlation of the three factors.

\section{Observations}
\label{obsec}

\subsection{Observations and Data Reduction} 

Observations of the M81 and NGC 2403 groups are summarized in \citet{chy08}, \citet{chy09}, and \citet{chy11}. 
We observed the CVn I, NGC 672, and NGC 45 groups in 32 sessions between January and August 2009.
We combined the observing sessions into 9 maps encompassing the group galaxies. Maps were made by moving the telescope in declination and sampling every 3$\arcmin$ at an integration time of 2-3 seconds per sample. Strips of constant declination were spaced by 3$\arcmin$. Corresponding maps were made by moving the telescope in right ascension to form a `basket weave' pattern over the region. Total integration time was approximately 58 hours for CVn I, 47 hours for NGC 672, and 41 hours for NGC 45.  Tables \ref{canes1obs} through \ref{n45obs} give a summary of the observations, including the RMS noise figures.

The GBT data were reduced in the standard manner using the GBTIDL and AIPS\footnote{Developed by the National Radio Astronomy Observatory;
documentation at http://gbtidl.sourceforge.net, http://www.aoc.nrao.edu/aips} 
data reduction packages.

In order to match our velocity resolution to the expected linewidths of HI 
clouds in the group, spectra were smoothed  
to a channel spacing of 24.4 kHz, corresponding to a velocity resolution of 
5.2 km s$^{-1}$. A reference spectrum for each of the observation sessions 
was made using an 
observation of an emission-free region, usually from the map edges. The reference spectrum was then used to 
perform a (signal-reference)/reference calibration of each pixel. 
The calibrated spectra were scaled by the system temperature, corrected 
for atmospheric opacity and GBT efficiency. We adopted the GBT efficiency equation 
(1) from \citet{lang07} with a zenith atmospheric opacity 
$\tau_{0}$ = 0.009. 

The frequency range observed was relatively free of RFI, with less than 0.5\% of 
all spectra adversely affected. The spectra exhibiting RFI were 
identified by tabulating the RMS noise level in channels free of neutral 
hydrogen emission.  Spectra that showed high RMS
noise across many channels were flagged and removed.  
Observations were gridded using the AIPS task SDIMG, which also averages 
polarizations. After amplitude calibration and gridding, a 1st-order polynomial 
was fit to line-free regions of the spectra and subtracted from 
the gridded spectra using the AIPS task IMLIN.  Only channels in high negative velocity ranges, where few or no sources are expected, were used for the fit. This simple baseline fit was extrapolated to the positive velocity range, 
where galaxies make a baseline fit unreliable. This baseline velocity range yielded a flat baseline for areas free of strong continuum radio sources, which is true of the majority 
of the region.
The effective angular resolution, determined 
from maps of 3C286, is 9.15\arcmin $\pm$ 0.05\arcmin. 
To convert to units of flux density, we observed the calibration source 3C286, 
whose flux density is 14.57 $\pm$ 0.94 Jy at 1.418 GHz  
\citep{ott94}. The calibration from 
K to Jy was derived by mapping 3C286 in the same way that the
 HI maps were produced. 
After all corrections for the GBT efficiency and the mapping process, the 
scale factor from K/Beam to Jy/Beam images is 0.43 $\pm$ 0.03.   Error estimates are difficult to obtain, given the wide range in RMS 
noise values over the maps. However, the dominant error 
contribution is the approximately 7\% uncertainty in the absolute calibration for these observations.

The mass detection thresholds were
calculated assuming that clouds would be unresolved in the GBT beam, using the
relation:

\begin{equation}
\left(\frac{\sigma_{M}}{M_{\odot}}\right)=2.36 \times 10^5 \left(\frac{D}{{\rm Mpc}}\right)^2 \left(\frac{\sigma_{s}}{{\rm Jy} }\right)
\left(\frac{\Delta V}{{\rm km}\hskip 1 mm {\rm s}^{-1}}\right) \sqrt N ,
\end{equation}

\noindent where $D$ is the distance in Mpc, $\sigma_s$ is the RMS noise in one channel, $\Delta V$
is the channel width, and $N$ is the number of channels required for a secure detection. We required a cloud candidate to be visible in at least $N=2$ channels, therefore the lowest velocity width that could be detected was 10 km s$^{-1}$ . 

Mass detection thresholds are listed in Tables \ref{canes1obs} through \ref{n45obs}.  HI cloud candidates were found by visual inspection of each channel of the baseline subtracted spectral line cube. Cloud candidates are spatially distinct from group galaxies and tidal streams and coherent in velocity over more than one channel. For each HI cloud candidate, we produced spectra (intensity versus velocity) using the AIPS task ISPEC. 

\section{Completeness of Study}
\label{complete}

\subsection{Velocity}

Our observations cover more than $\pm$ 700 km s$^{-1}$ with respect to the systemic velocity
of any group galaxy. In principle, this allows us to detect clouds associated with any of the
mechanisms that create HI clouds, all of which predict clouds to be within 300 km s$^{-1}$ of the galaxy's velocity \citep{miller09}.

\subsection{Position} 

Studies of the Milky Way and M31 find clouds anywhere within 1-50 kpc of a
galaxy center \citep{wvw97,thilker04}. N-body simulations of galaxy interactions predict that clouds and large-scale structure will be found as close as 3 kpc and as far as 750 kpc from galaxies \citep{miller09}. In order to balance the requirement of a good mass sensitivity and a large angular area in our observations, we designed our observations to cover a projected radius of at least 40-50 kpc around major group galaxies. 
The map sizes are adequate to detect both interaction remnants and DM halo-associated clouds.
The angular resolution of our observations ranges from 10-20 kpc. Since galactic fountain HI clouds are found at a distance of $<$ 10 kpc from the galactic disk, we may not be able to resolve galactic fountain clouds, depending on their velocity and orientation. 


\subsection{Mass}

The 5$\sigma$ detection threshold for HI masses ranged from 9.7 $\times$ 10$^5$ to 12.7 $\times$ 10$^6$  M$_{\odot}$,
assuming a linewidth of 10 km s$^{-1}$ and using the distance of the most massive group galaxy.  With these mass detection thresholds, we would be able to detect analogs of the most massive HI clouds that have been detected surrounding nearby galaxies, as well as large Milky Way objects such as Complexes C and H and the Magellanic Stream. These clouds are the most likely to originate in galaxy interactions, cold accretion, and dark matter halos, rather than from supernovae \citep{miller09}, so our study is biased towards HI clouds with an extragalactic origin. To estimate the completeness of the survey in terms of mass, we compiled the published values for all known extraplanar HI clouds surrounding nearby galaxies and the Milky Way. If this mass spectrum is representative of the general extragalactic HI cloud mass function, we can detect anywhere from 12\% to 35\% of the analogous clouds surrounding the galaxies in our sample. Table \ref{masscomptable} lists the average mass detection threshold and mass completeness for each galaxy group. Figure \ref{massdist} shows the mass function of known HI clouds compared to our observational limits for each galaxy group. Note that the mass function for HI clouds is not well-constrained, particularly for the Milky Way, since the properties of Milky Way HI clouds beyond the stellar halo are not known. We should be able to detect cold accretion clouds,  which are predicted to have HI masses of 10$^6$ to 10$^7$ M$_{\odot}$  \citep{keres09}.

\section{Predictions of Numerical Simulations}
\label{simulations}

In order to determine properties expected for gas-embedded dark-matter halos, we analyzed several N-body simulations.  
We conducted two dark matter-only cosmological N-body simulations for comparison. One simulation uses initial conditions from WMAP3, and one from WMAP5. We simulated a 50$^3$ (Mpc/h)$^3$ volume of the universe from z=149 to z=0 using 256$^3$
particles. At redshift zero, we selected a
volume 10 Mpc/h on a side to resimulate with 512$^3$ particles with a
'zoom' technique aimed to preserve the tidal field outside the higher
resolution volume. This smaller volume was selected to host a
$\sim 10^{12} M_\odot$ halo at z=0, consistent with the M81
galaxy halo.   We used a friends-of-friends algorithm with a linking length of $b=0.2$  to
identify dark matter halos \citep{davis85}, and used a SUBFIND technique to determine the bound subhalos \citep{springel01}.

We also use the dark matter-only ``set C'' simulations from \citet{emil} in which a 65$^3$ (Mpc/h)$^3$ box was simulated with 408$^3$ particles and WMAP3 cosmological parameters. Nine dark matter halos with virial masses at z=0 of 1.6-2.5 $\times$ 10$^{12}$ M$_{\odot}$ were resimulated using a zoom technique with mass resolution similar to the simulations described above. The AHF halo finder \citep{kno2009} was used to determine the bound subhalos. 

In order to connect the numerical simulations with the properties of HI observations, 
we require an estimate of the HI to dark matter mass fraction. The distribution of HI gas compared to the dark matter halo distribution is uncertain. For the purpose of comparing our HI observations to the simulated HI clouds, we assume that HI gas follows the dark matter distribution. We include in our comparison the prescription of \citet{gne00}, which includes a mass cutoff such that a dark matter halo with mass less than 2 $\times$ 10$^8$ M$_\odot$ has no associated HI gas. We take as the scaling factor from dark matter to HI gas mass 
the \citet{zwaan03} results, summarized in \citet{fuku04}. 
Their values yield a HI to dark matter 
mass fraction of 0.0018 $\pm$ 0.0003. 

\begin{figure}
\centering
\begin{tabular}{cc}
\includegraphics[width=3.5in]{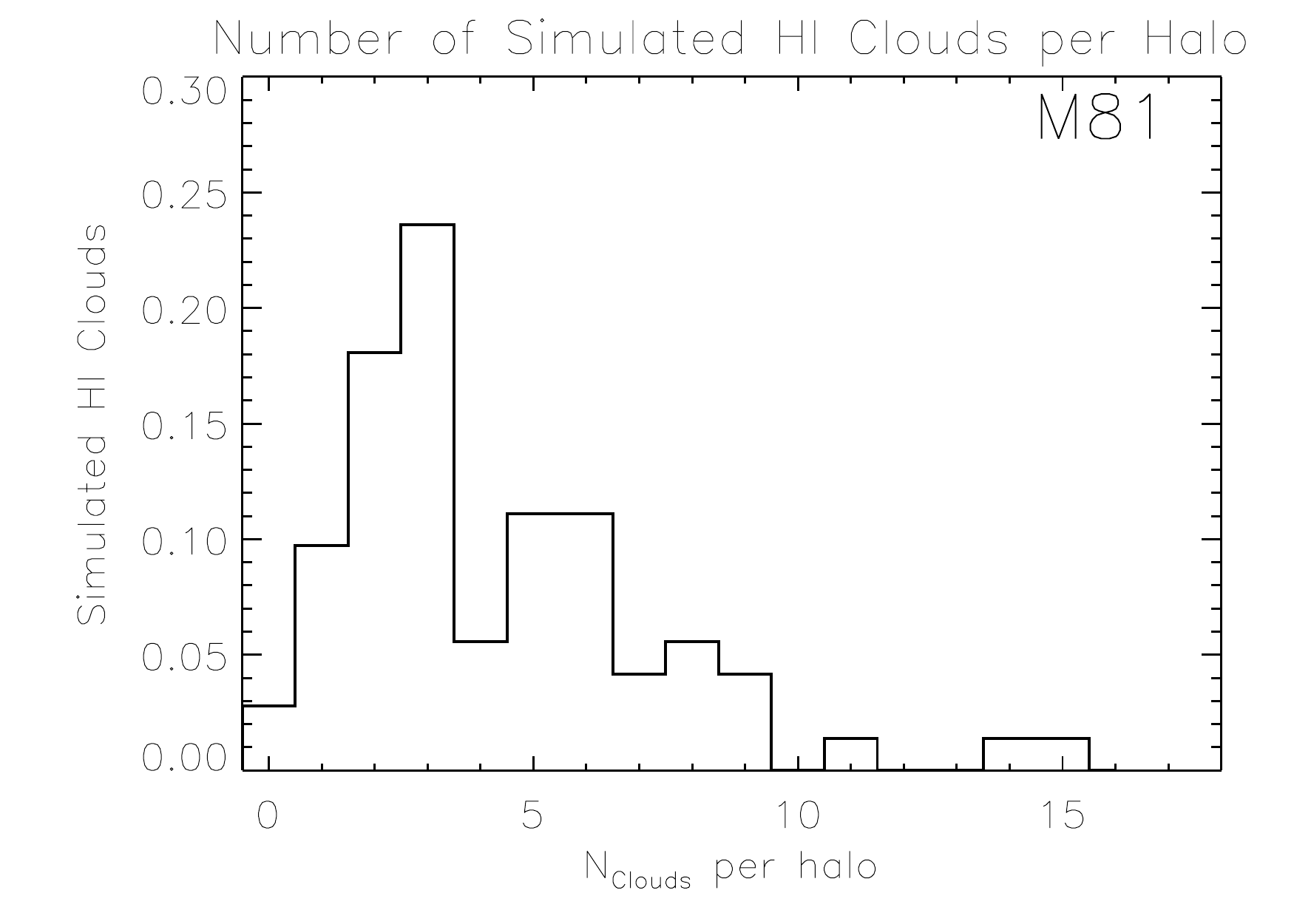} &
\includegraphics[width=3.5in]{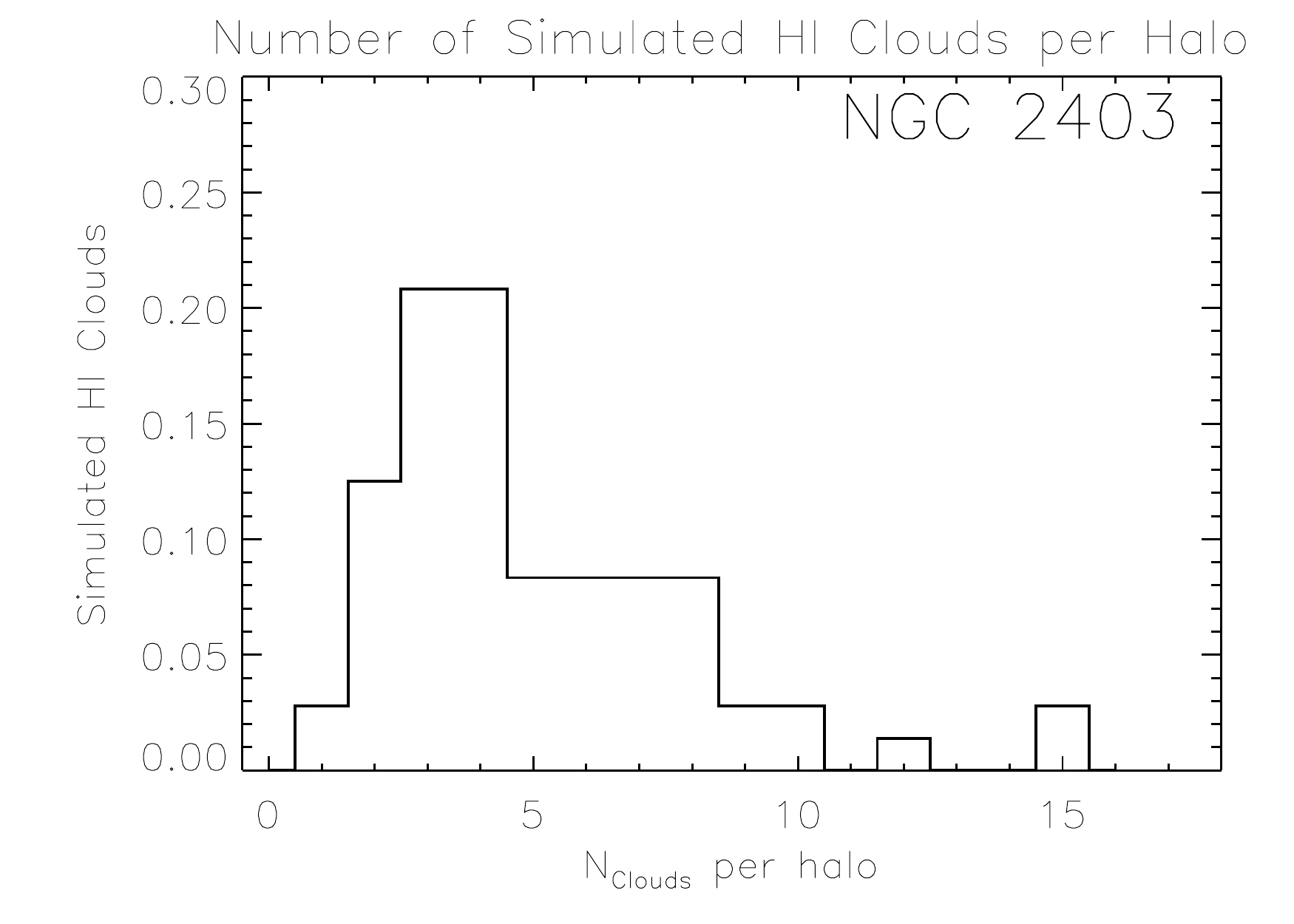} \\
\includegraphics[width=3.5in]{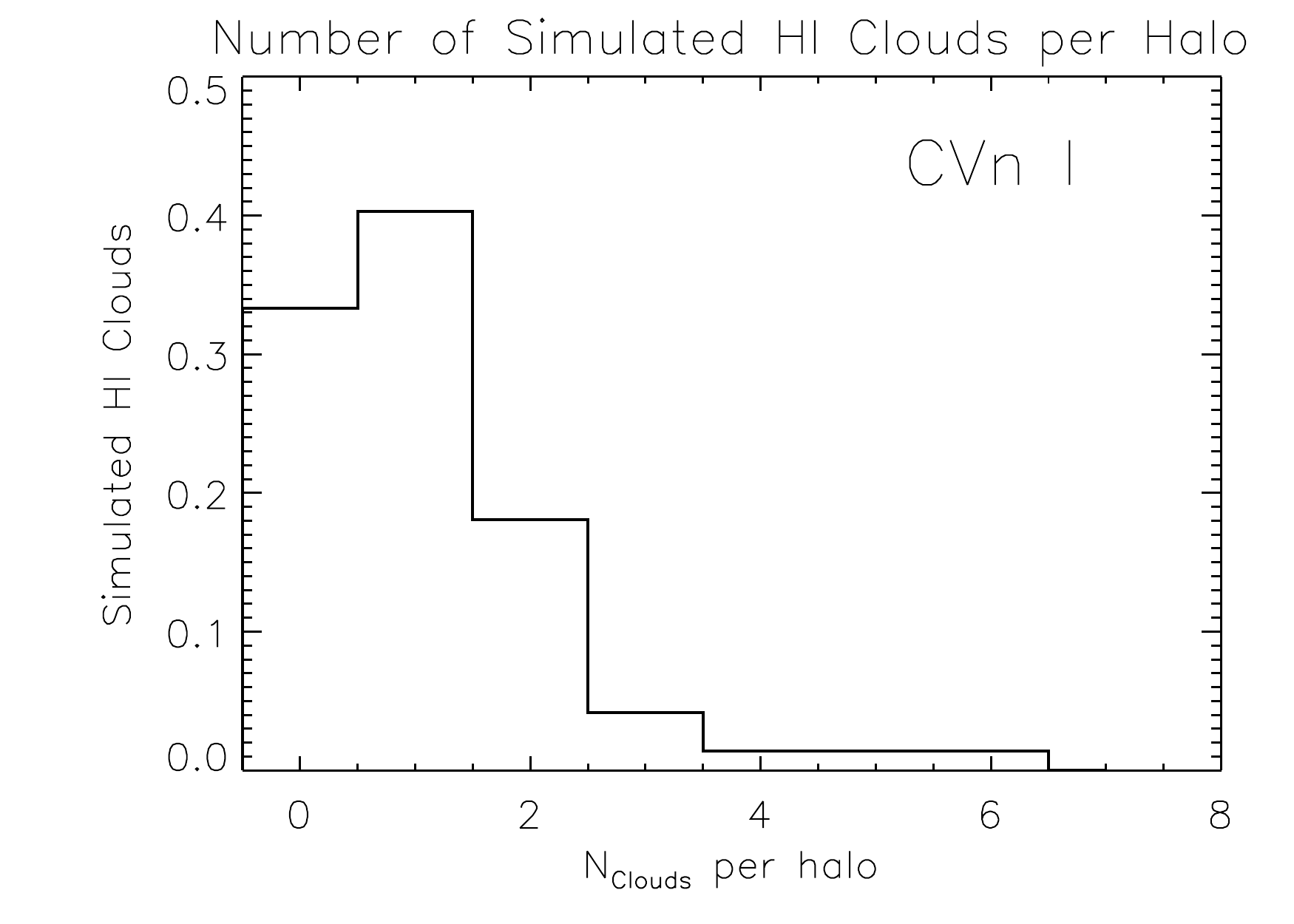} &
\includegraphics[width=3.5in]{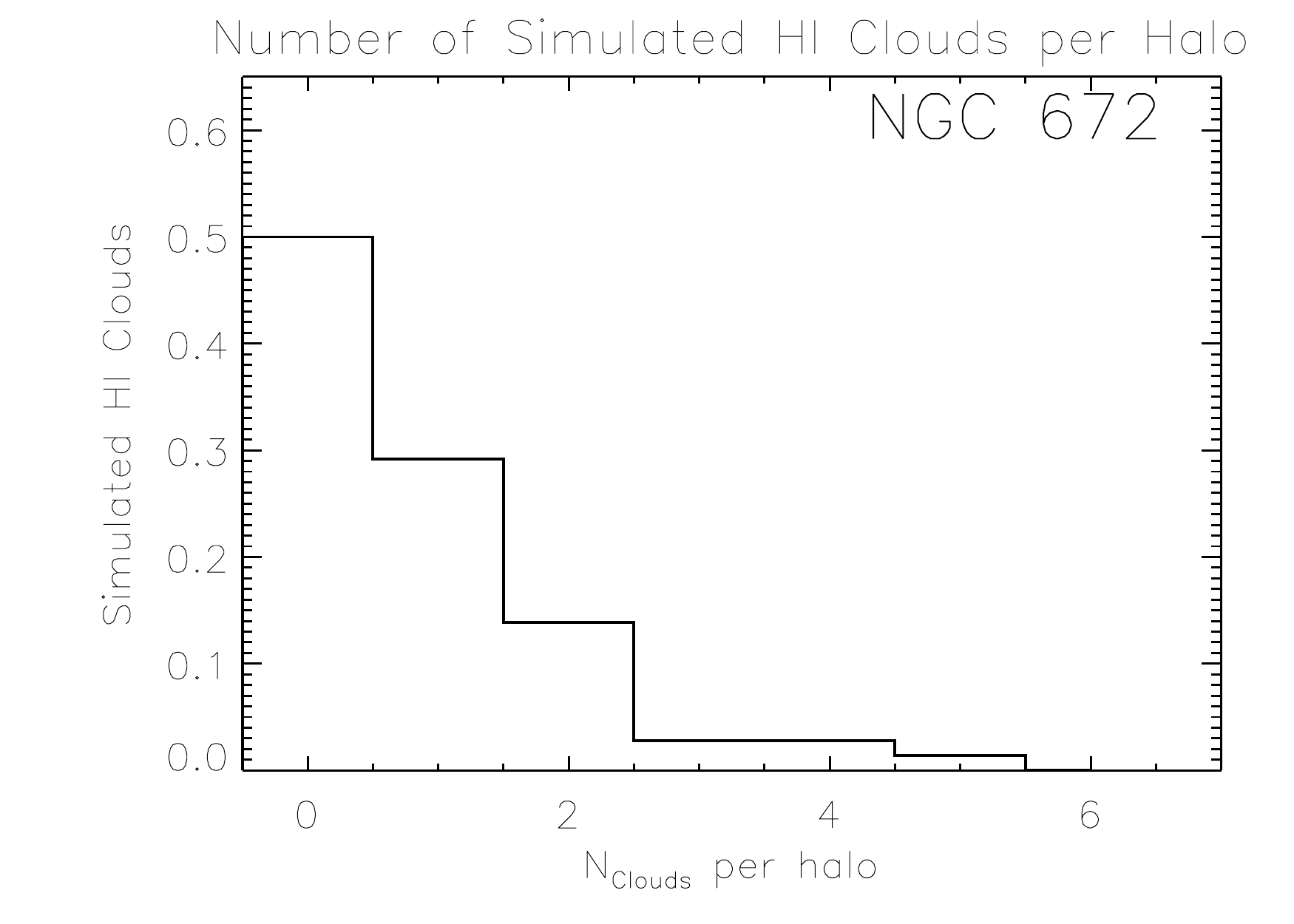} \\
\includegraphics[width=3.5in]{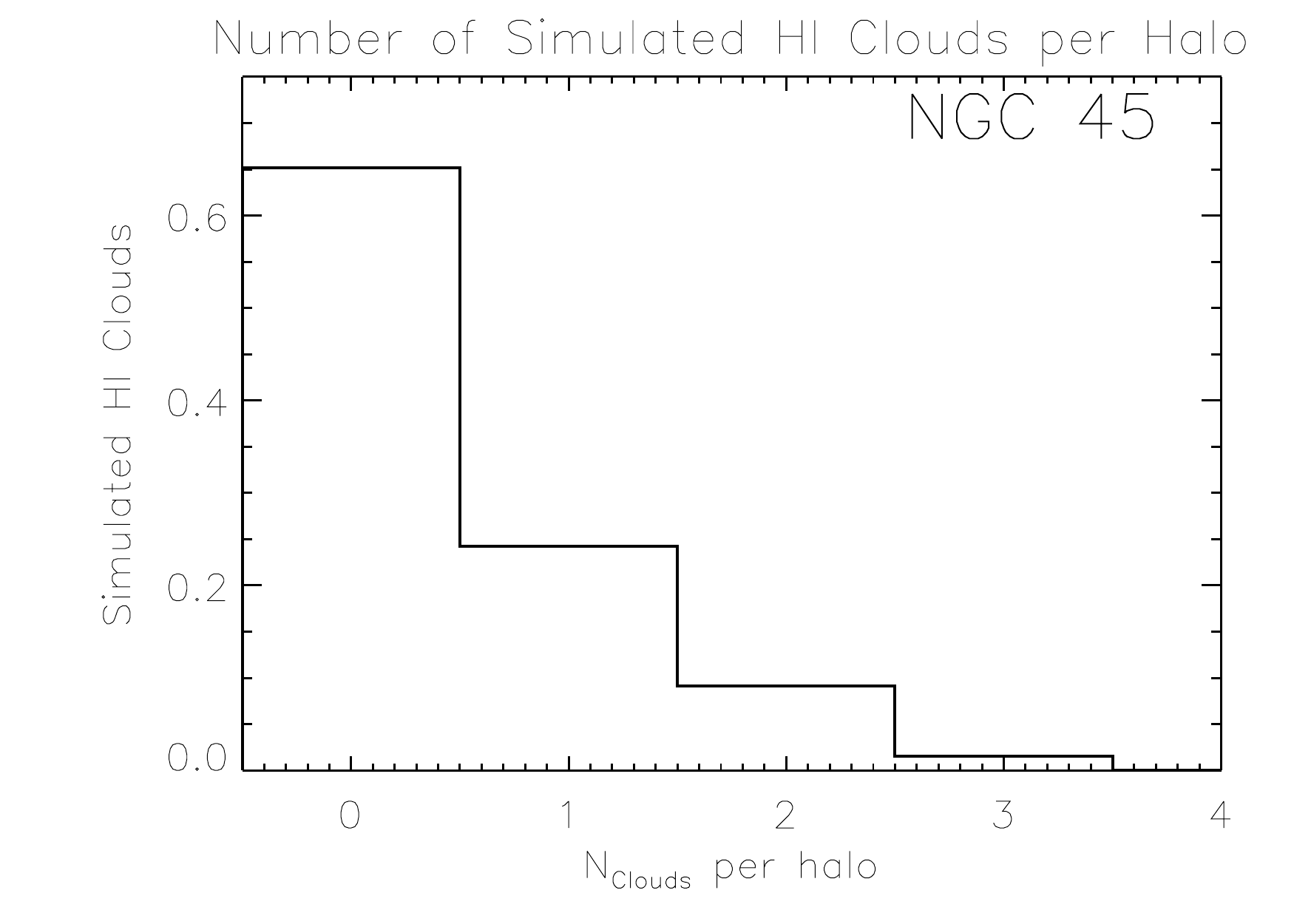} & \\
\end{tabular}
\caption{Number distribution function of simulated clouds over all simulations described in Section \ref{simulations}. Diagrams include all dark matter halos above the mass detection threshold for the galaxy group (assuming a HI to dark matter 
mass fraction of 0.0018 $\pm$ 0.0003 as described in Section \ref{simulations}), that are within $\pm$ 700 km  s$^{-1}$ and $\pm$ 50 kpc of the most massive galaxy in each galaxy group. The distributions plotted here show the number of subhalos per major halo. The distribution functions are normalized so that their sums are unity.}
\label{sim_num}
\end{figure}

\begin{figure}
\centering
\begin{tabular}{cc}
\includegraphics[width=3.5in]{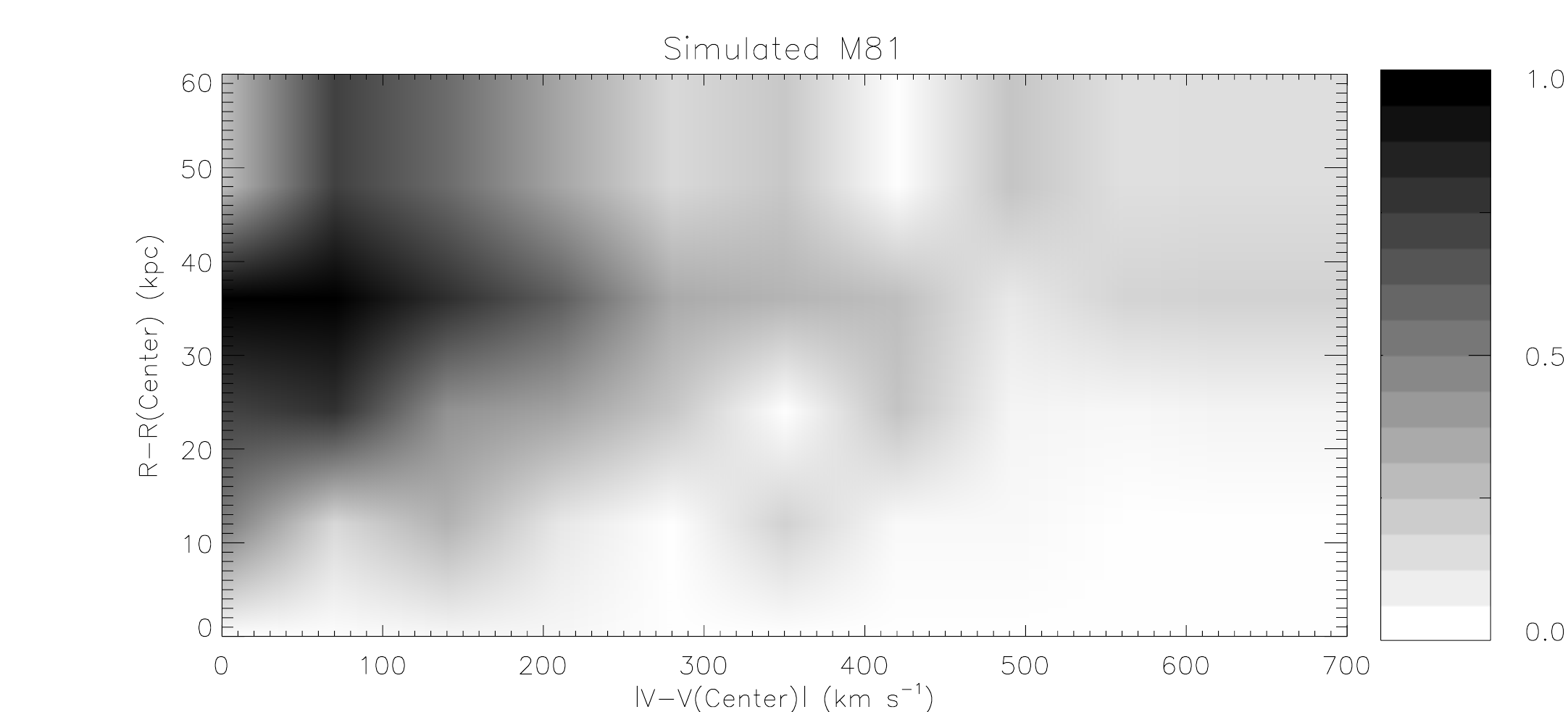} &
\includegraphics[width=3.5in]{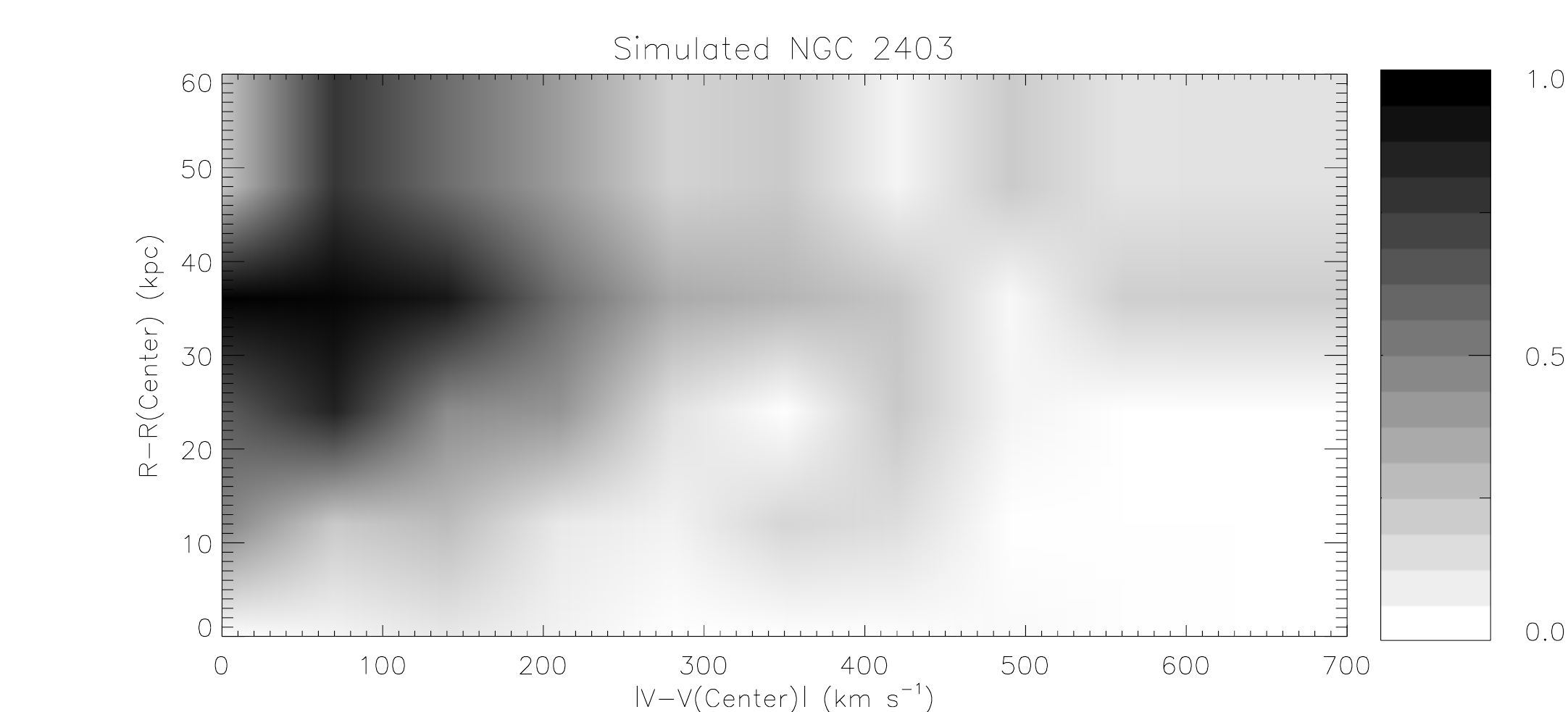} \\
\includegraphics[width=3.5in]{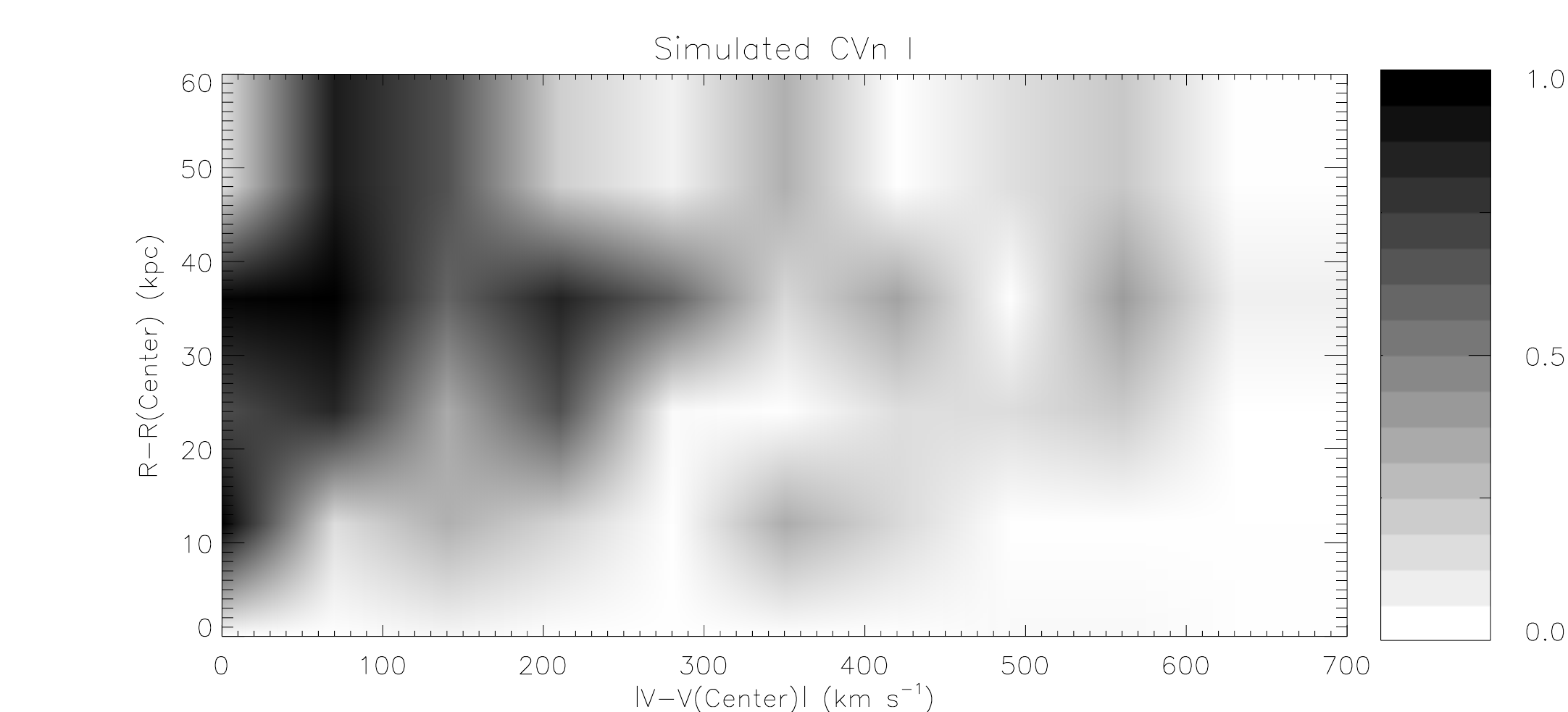} &
\includegraphics[width=3.5in]{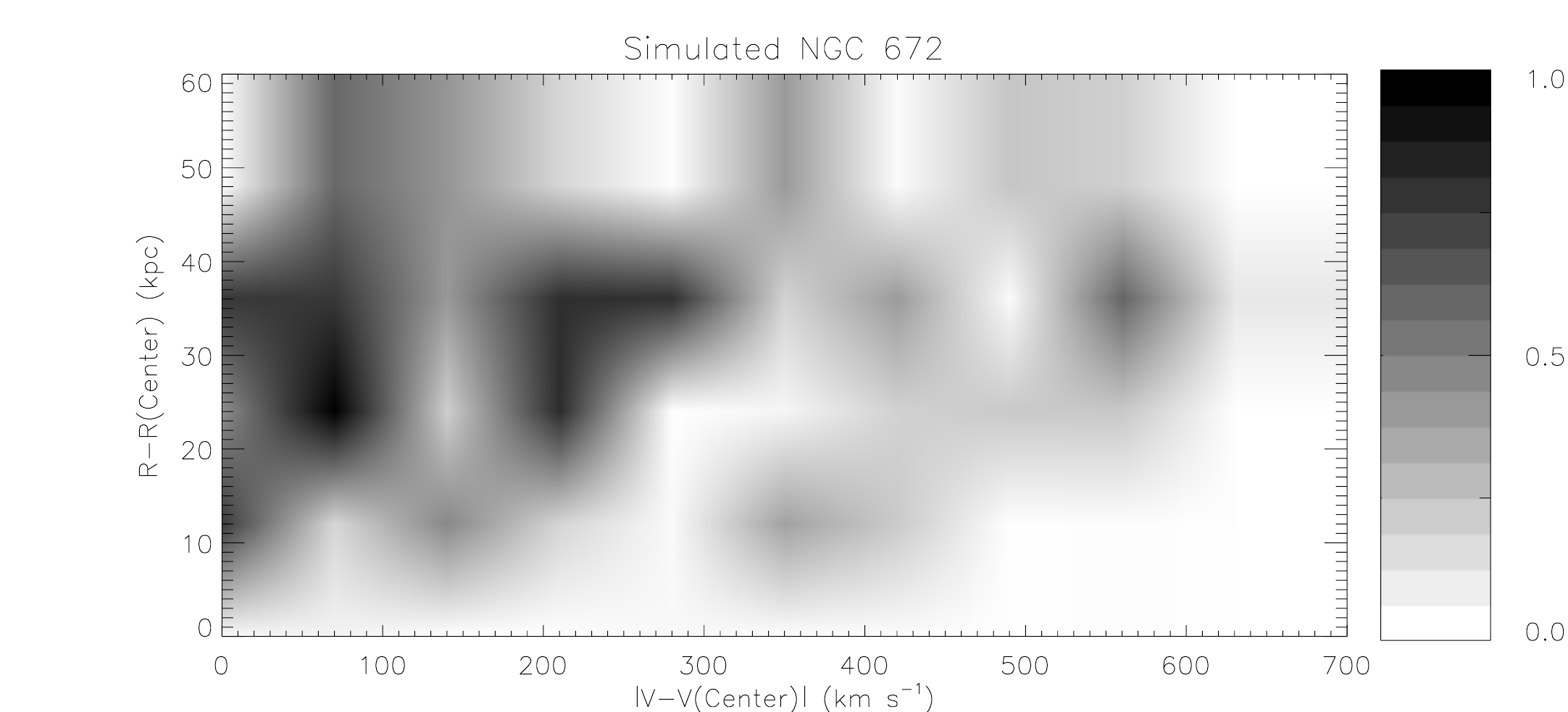} \\
\includegraphics[width=3.5in]{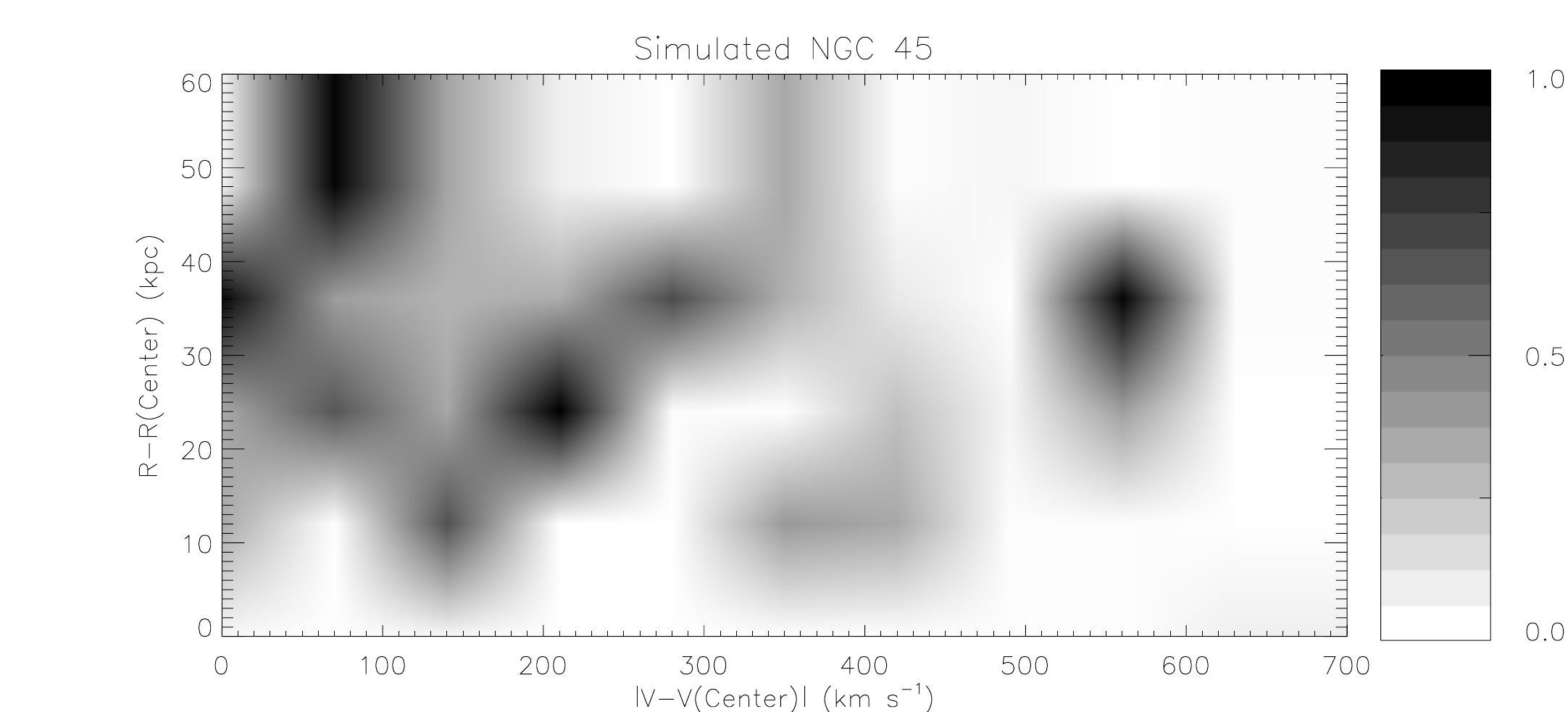} & \\
\end{tabular}
\caption{Position-velocity distributions of simulated clouds.  Distributions include all simulated dark matter halos above the mass detection threshold for the galaxy group (assuming a HI to dark matter 
mass fraction of 0.0018 $\pm$ 0.0003 as described in Section \ref{simulations}), that are within $\pm$ 700 km  s$^{-1}$ and $\pm$ 50 kpc of the most massive galaxy in each galaxy group. The greyscale for each plot runs from zero to the maximum number of subhalos detected in that simulated galaxy group, normalized so that the maximum is unity.}
\label{sim_phase}
\end{figure}

\begin{figure}
\centering
\begin{tabular}{cc}
\includegraphics[width=3.5in]{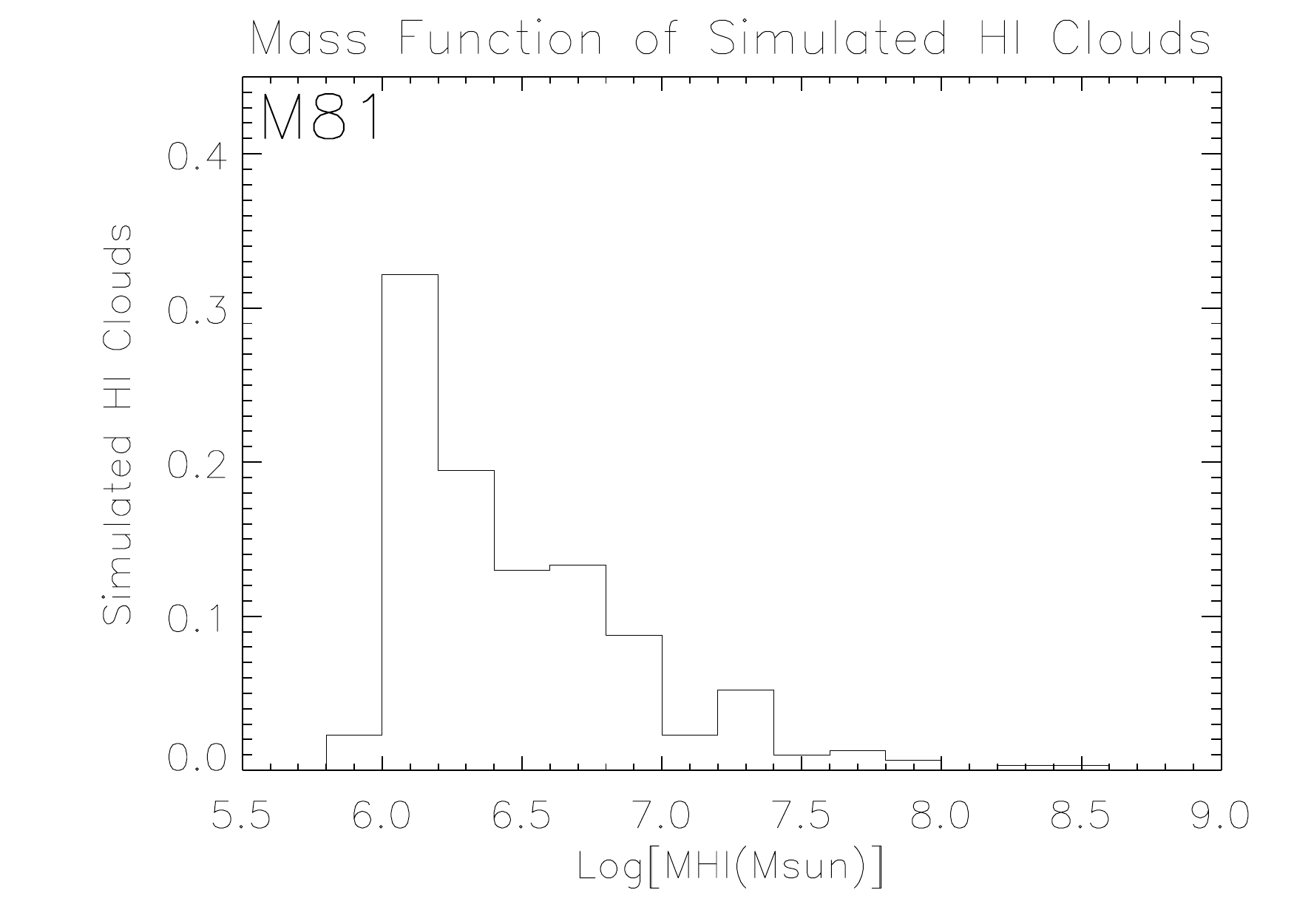} &
\includegraphics[width=3.5in]{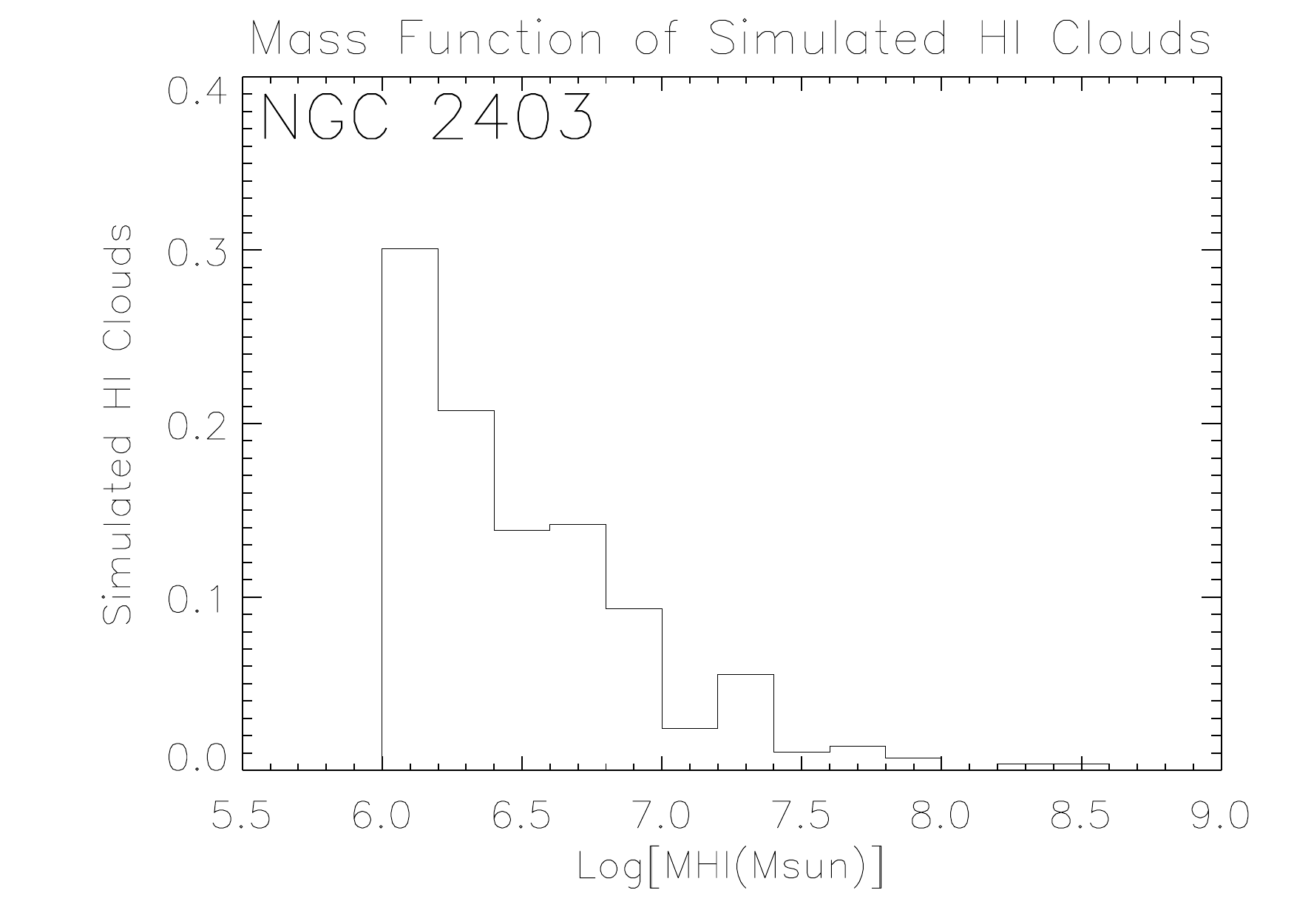} \\
\includegraphics[width=3.5in]{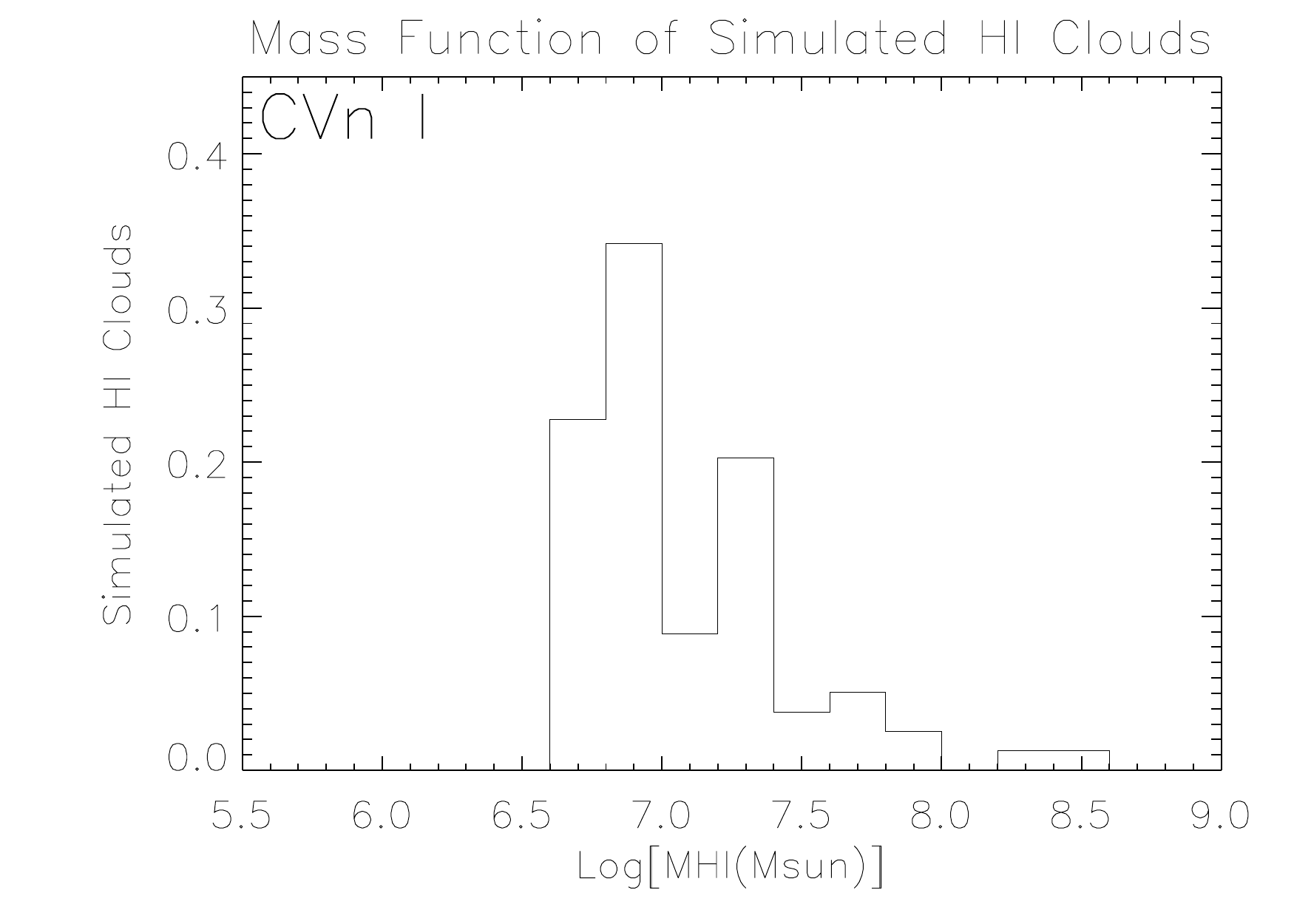} &
\includegraphics[width=3.5in]{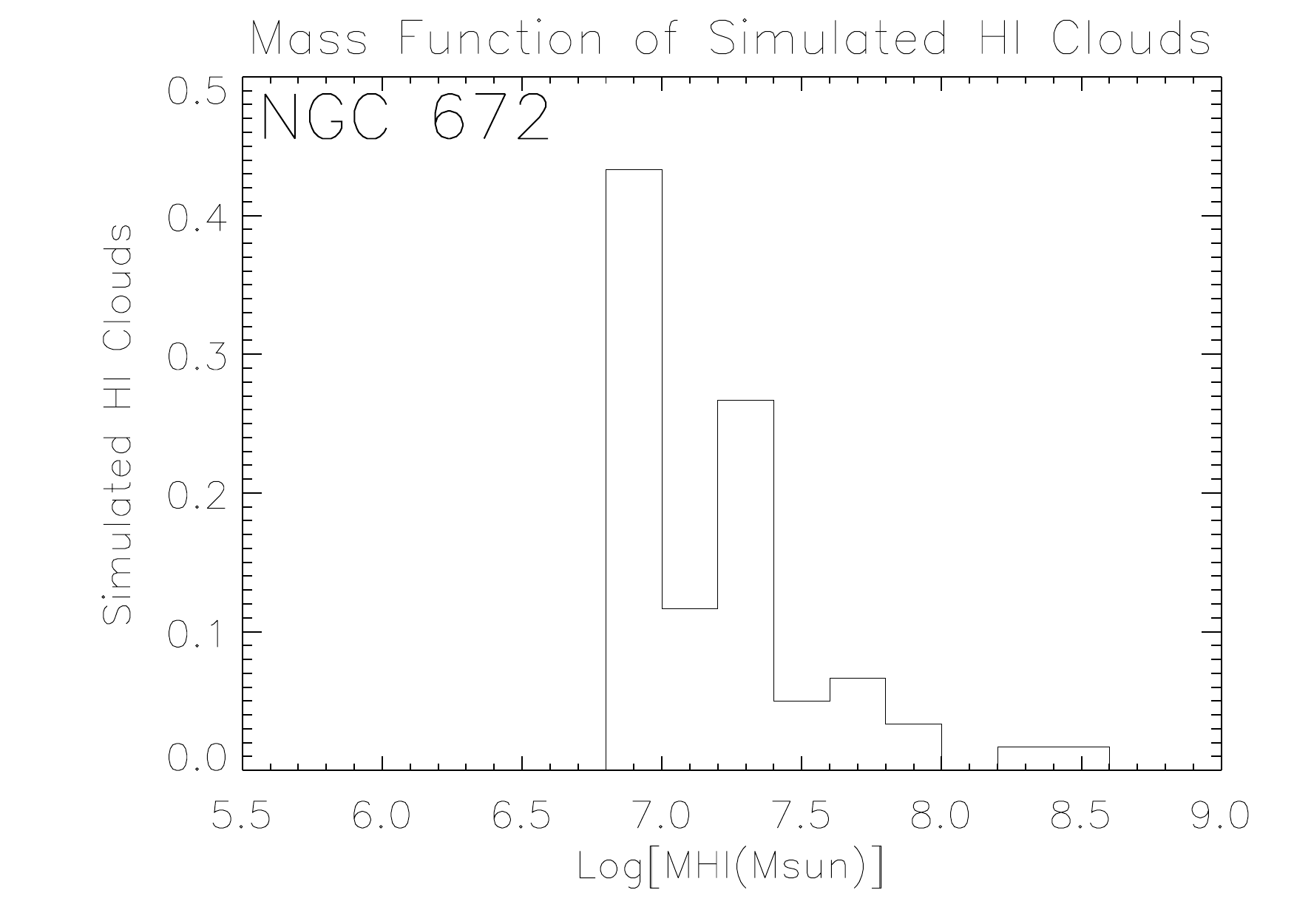} \\
\includegraphics[width=3.5in]{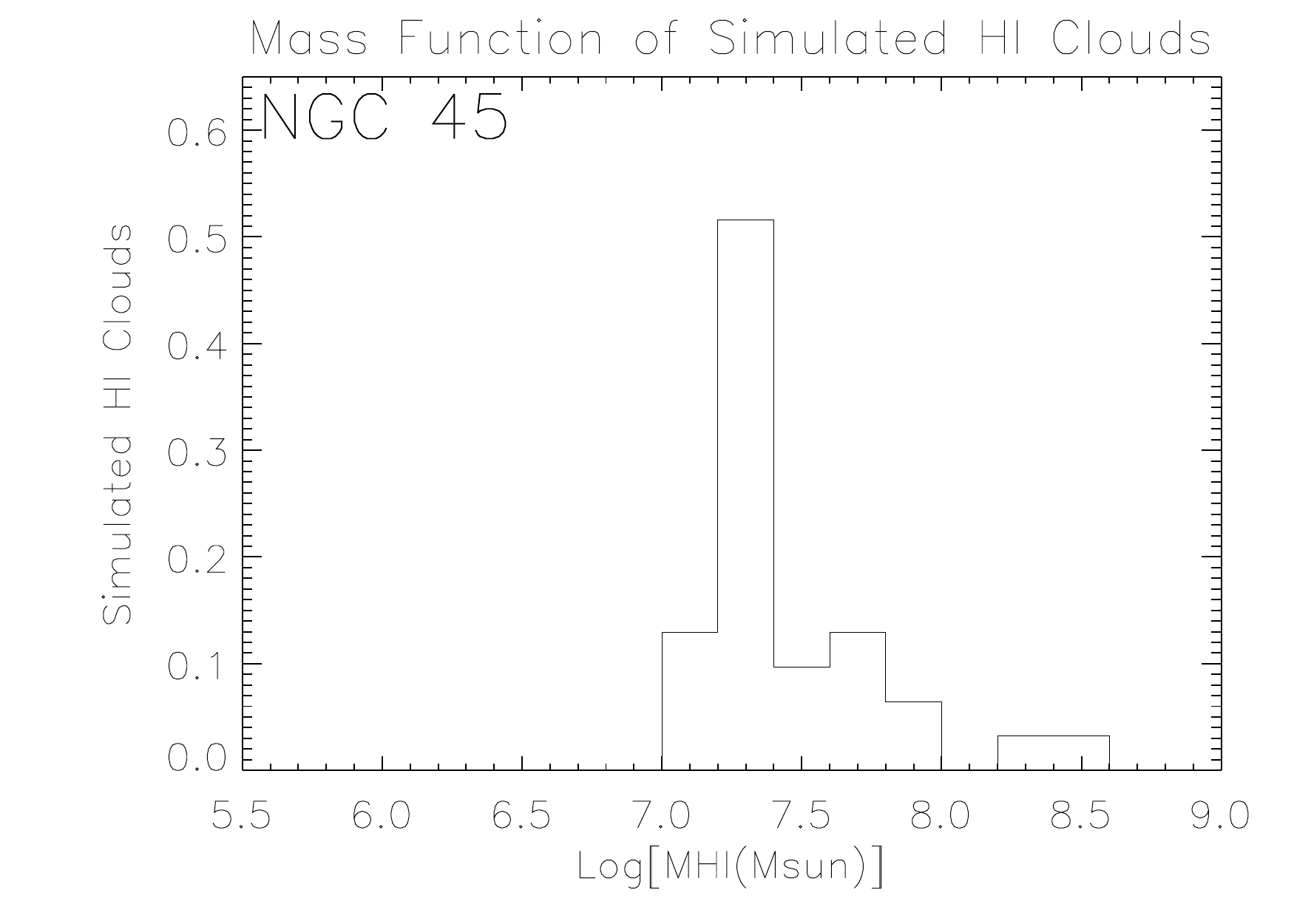} & \\
\end{tabular}
\caption{HI mass functions of simulated clouds. Distributions include all simulated dark matter halos above the mass detection threshold for the galaxy group (assuming a HI to dark matter 
mass fraction of 0.0018 $\pm$ 0.0003 as described in Section \ref{simulations}), that are within $\pm$ 700 km  s$^{-1}$ and $\pm$ 50 kpc of the most massive galaxy in each galaxy group. The mass distribution functions are normalized so that their sums are unity.}
\label{sim_mass}
\end{figure}

After scaling the simulated dark matter to HI, we apply the 5$\sigma$ HI mass threshold, velocity limits, and angular size limits of our survey. 
For each simulation, we calculate the probability that detectable sub-halos reside around the major dark-matter halos. This probability depends on the mass sensitivity of each observation, ranging from 38\% of major dark-matter halos hosting detectable sub-halos at the mass sensitivity of the NGC 45 group observation, to 97\% at the M81 and NGC 2403 groups mass sensitivity. For each major dark matter halo with sub-halos that would be detectable as HI clouds in our observations, we find the properties of those sub-halos.  We repeat this process rotating each simulation box three ways.   In the simulations, we find between 0 and 16 clouds with an average of 3 clouds of sufficient mass to be detected in our observations. Figure \ref{sim_num} shows the distribution of number of detectable simulated HI clouds per major dark-matter halo. It is important to note that in the NGC 672 and NGC 45 groups, the probability of the major halo hosting a dark-matter associated HI cloud detectable by our observations is $\leq$ 50\%. Figure \ref{sim_phase} shows the position-velocity distributions of the simulated HI clouds for each galaxy group, and Figure \ref{sim_mass} shows the simulated HI mass functions. The simulated clouds have a wide distribution of position and velocity separations from the central galaxy, ranging out to the very edge of our observations.

Table \ref{simtable} presents the detailed predictions of the simulations. It is worth noting that the predicted properties (positions and velocities) of the clouds are remarkably similar, regardless of the varied mass sensitivities of the observations. The sensitivity of the observation appears to affect only the number of clouds we should detect, and not their properties. That is, the positions and velocities of the simulated HI clouds do not correlate with their masses. Therefore, we would not expect that more sensitive observations would find HI clouds with similar position and velocity distributions to those observed in the M81 group HI clouds. No HI clouds were detected in the corresponding region of our observations of the CVn I, NGC 672, and NGC 45 groups.

The above analysis does not account for numerical destruction effects in the simulations. Due to the gravitational softening lengths employed in N-body simulations subhalos orbiting within the halo of a larger galaxy experience extraneous tidal fields than can artificially disrupt subhalos. These effects are dominant for poorly resolved subhalos in the inner halo region that have short dynamical times. A method for estimating the amount of numerical destruction in simulations is presented in \citep{emil} and applied to their 9 set C halos the method estimates that between 2 to 6 subhalos with dark matter masses $> 2 \times 10^8 M_{\odot}$ were numerically destroyed within 50 kpc of each halo center. This would increase the predicted number of detectable subhalos and make the disparity with the observations more striking.

However, our dark matter-only simulations do not include the effects of baryons on the dark matter subhalo abundances. Dark matter subhalos whose orbits intersect a baryonic disk can be destroyed by disk shocking \citep{OSC72}. Calculations show a reduction in the dark matter subhalo abundances by factors of $2-3$ for masses $10^9 - 10^7 M_{\odot}$ in the inner $\sim 30$ kpc ($\sim 5$ disk scale-lengths) of a Milky Way-sized galaxy with a disk \citep{DSHK10}. Disk shocking is likely to offset the gains in subhalo numbers from considerations of numerical destruction for group galaxies with disk structures. 

In \citet{chy11}, we presented observations of the M81 and NGC 2403 groups. We found 9 HI clouds associated with the M81 group so in this case a statistical comparison was possible. \citet{chy11} presents a detailed description of the clouds. Table \ref{fil_cloudprop} summarizes the properties of the clouds in the M81 group for reference. To summarize, the detected HI clouds have masses of 10$^{6-8}$ M$_{\odot}$, and have a mean velocity deviation of 125 km s$^{-1}$ and mean projected distance of 110 kpc from M81. In contrast, analysis of numerical simulations (in the same manner described in this section) predicts that we should detect 40-50 HI clouds embedded within dark matter minihalos. These simulated clouds would have masses of 10$^{6-8}$ M$_{\odot}$, average velocity deviation from the simulated ``M81" galaxy of 250 km s$^{-1}$, and mean projected separation of 170 kpc.  Although the detected HI cloud mean masses match the simulated mean masses, the observed and simulated mass functions are discrepant. A KS-test of the two mass functions yields only a 0.3\% probability that the mass functions are statistically similar. Additionally, the number count and distributions of positions and velocities of detected HI clouds are statistically inconsistent with the distributions expected for dark matter halos with embedded HI based on the simulations \citep{chy11}. For the groups observed for this paper, no HI clouds were detected so the discrepancy between predictions and observations
is further increased.

\section{Results, Conclusions and Future Work}
\label{consec}

We have observed 5 nearby galaxy groups with the GBT in an effort to search for HI clouds associated with either galaxy interactions or dark-matter halos. 
In \citet{chy11}, we presented observations of the M81 and NGC 2403 groups. We found 9 HI clouds associated with the M81 group, and no clouds in the NGC 2403 group. In this study, we observed the CVn I, NGC 672 and NGC 45 groups.  The 5 observed groups have varying levels of galaxy interactions, which we quantify using an interaction index.

\begin{figure}
\centering
\includegraphics[height=5.75in]{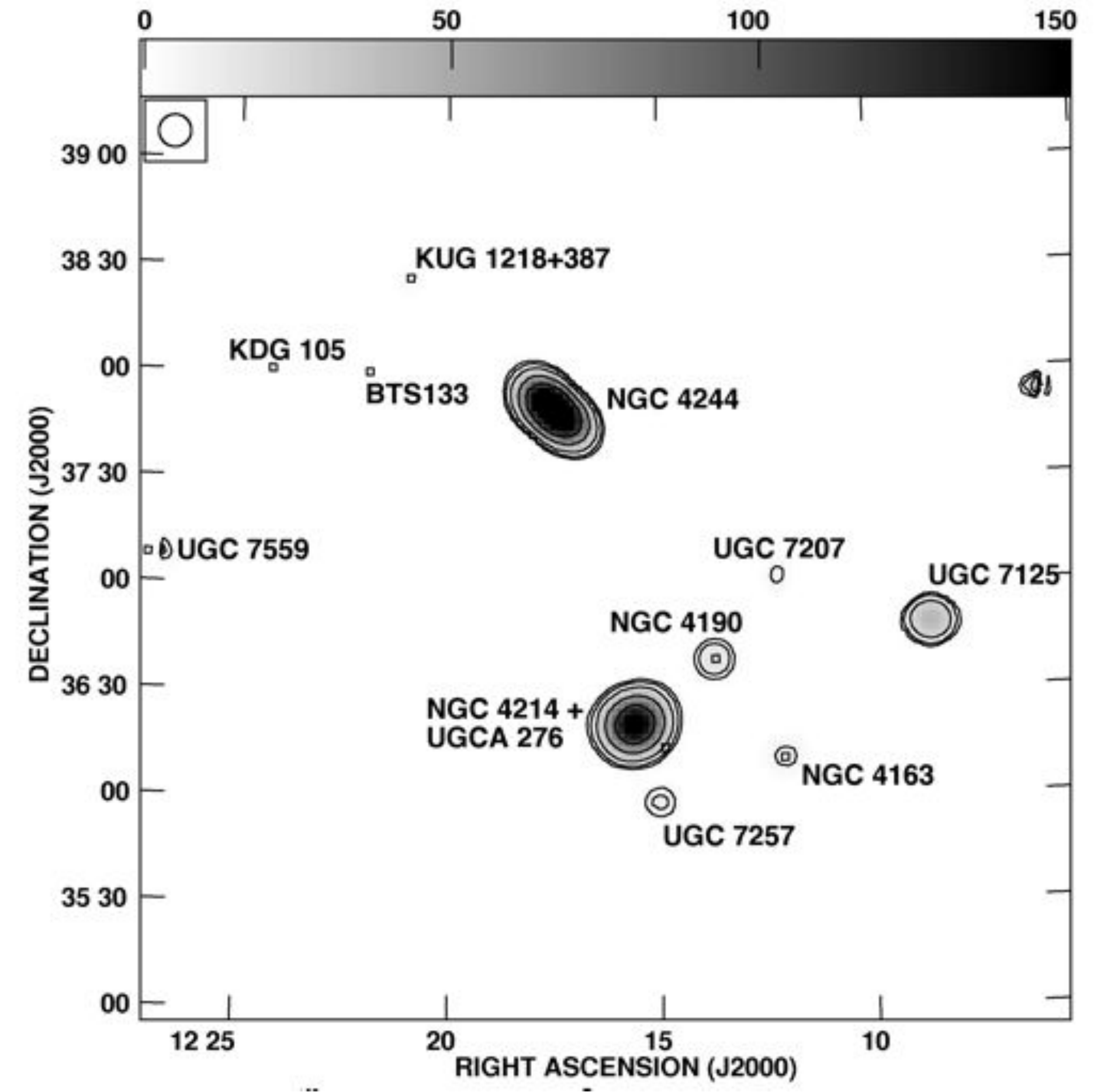}
\caption{\textbf{CVn I Group HI column density I}:  Group galaxies and background galaxies are labeled. All flux above 1$\sigma$ is included. Grayscale: 0-150 Jy beam$^{-1}$ $\times$ km  s$^{-1}$. Contours: 2 Jy beam$^{-1}$ $\times$ km  s$^{-1}$ $\times$(-3,3,5,10,25,50,100,250)}
\label{mom0_canes1}
\end{figure}

\begin{figure}
\centering
\begin{tabular}{cc}
\includegraphics[height=2.75in]{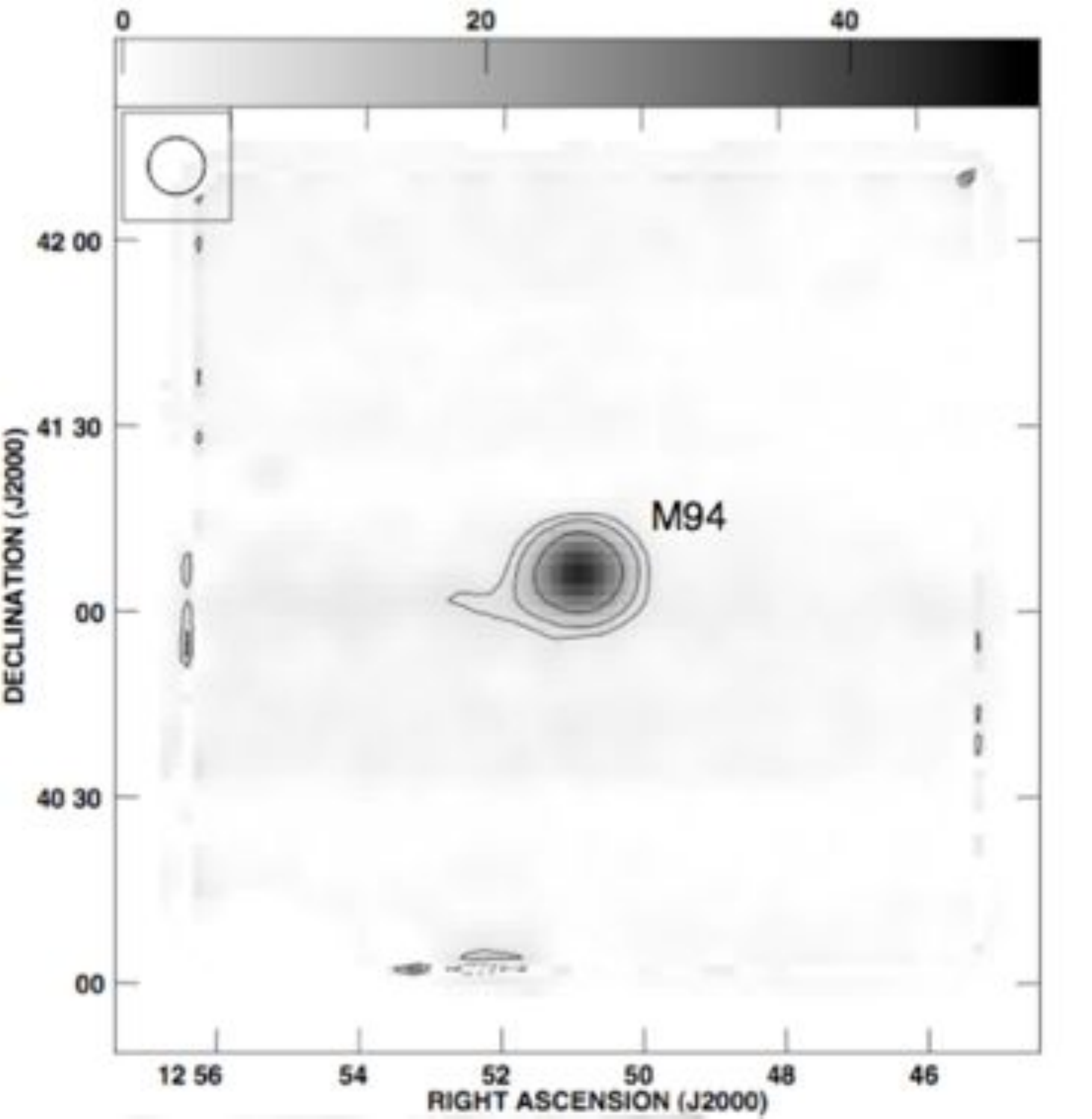} & 
\includegraphics[height=2.75in]{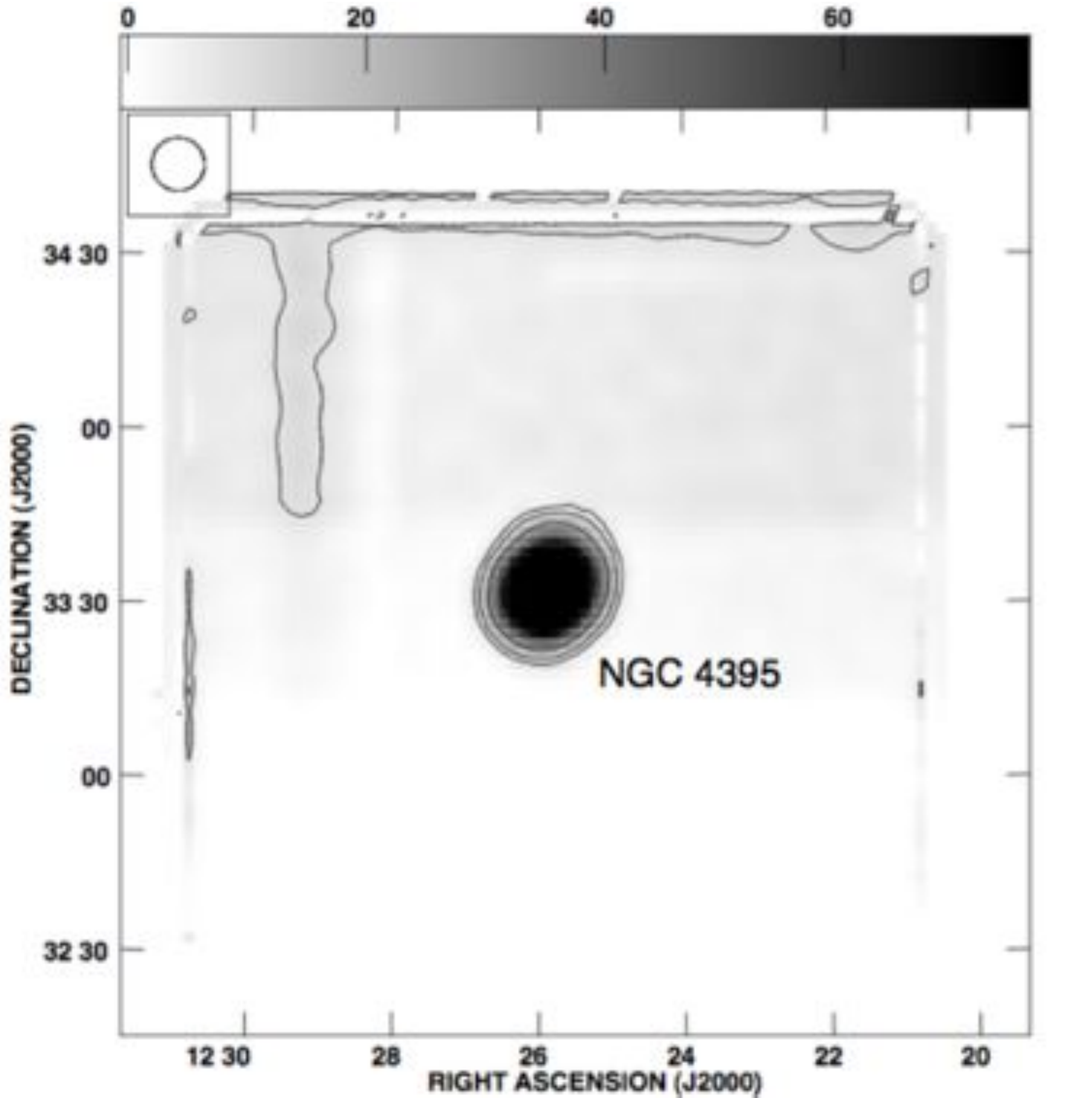} \\
\end{tabular}
\caption{\textbf{CVn I group column density maps II and III.} Left: HI column density map of M94. Grayscale: 0-50 Jy beam$^{-1}$ $\times$ km  s$^{-1}$. Contours: 2 Jy beam$^{-1}$ $\times$ km  s$^{-1}$ $\times$(-3,3,5,10,25,50,100,250. Right: HI column density map of NGC 4395. Grayscale: 0-75 Jy beam$^{-1}$ $\times$ km  s$^{-1}$. Contours: 2.75 Jy beam$^{-1}$ $\times$ km  s$^{-1}$ $\times$(-3,3,5,10,25,50,100,250)}
\label{m94_4395}
\end{figure}

\begin{figure}
\centering
\includegraphics[height=5in]{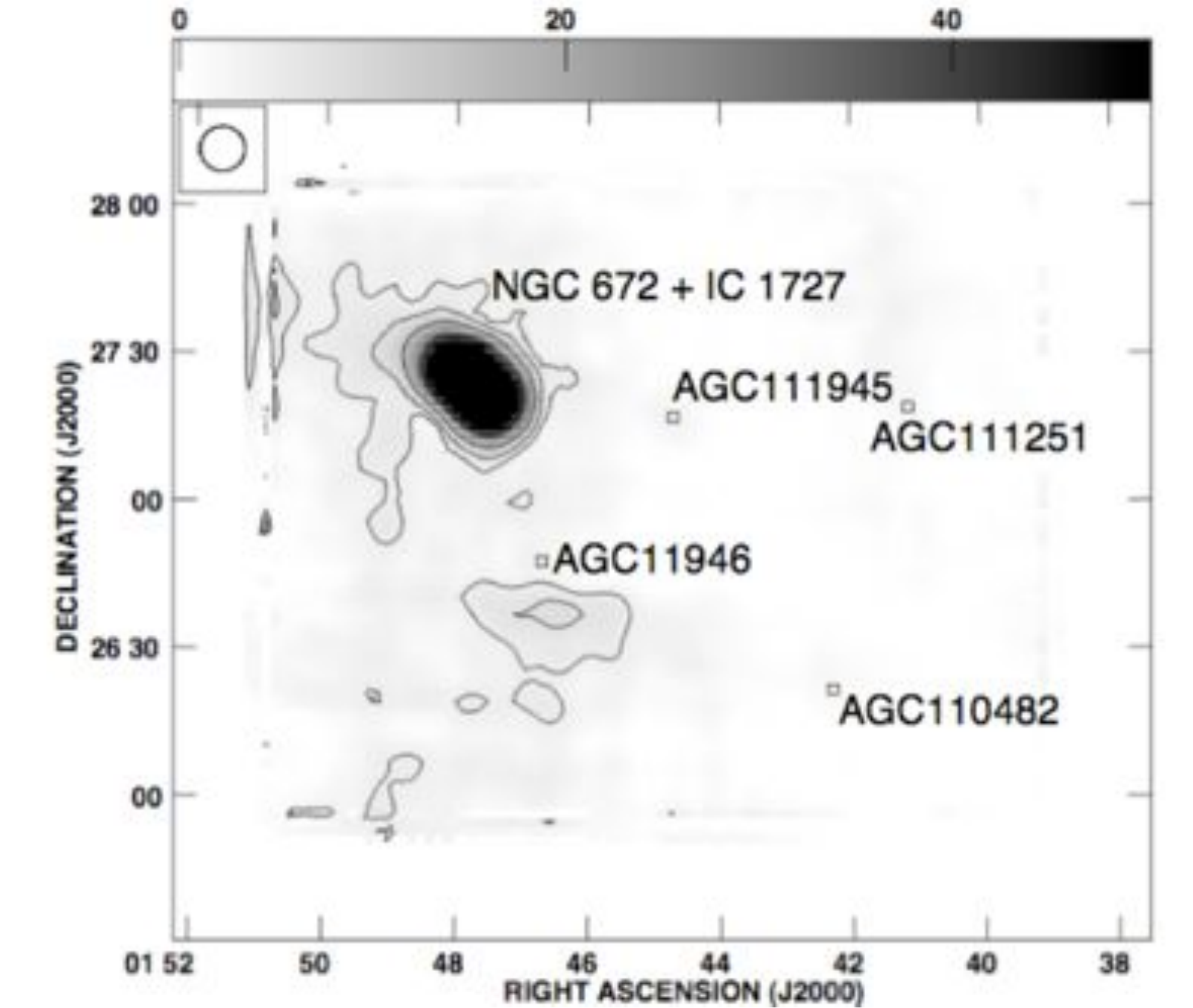}
\caption{\textbf{NGC 672 Group HI column density I:}  Any emission not associated with labeled galaxies is likely due to noise, since it does not have coherent velocity structure. Grayscale: 0-115 Jy beam$^{-1}$ $\times$ km  s$^{-1}$. Contours: 3.25 Jy beam$^{-1}$ $\times$ km  s$^{-1}$ $\times$(-3,3,5,10,25,50,100,250)}
\label{mom0_n672}
\end{figure}

\begin{figure}
\centering
\begin{tabular}{cc}
\includegraphics[height=2.75in]{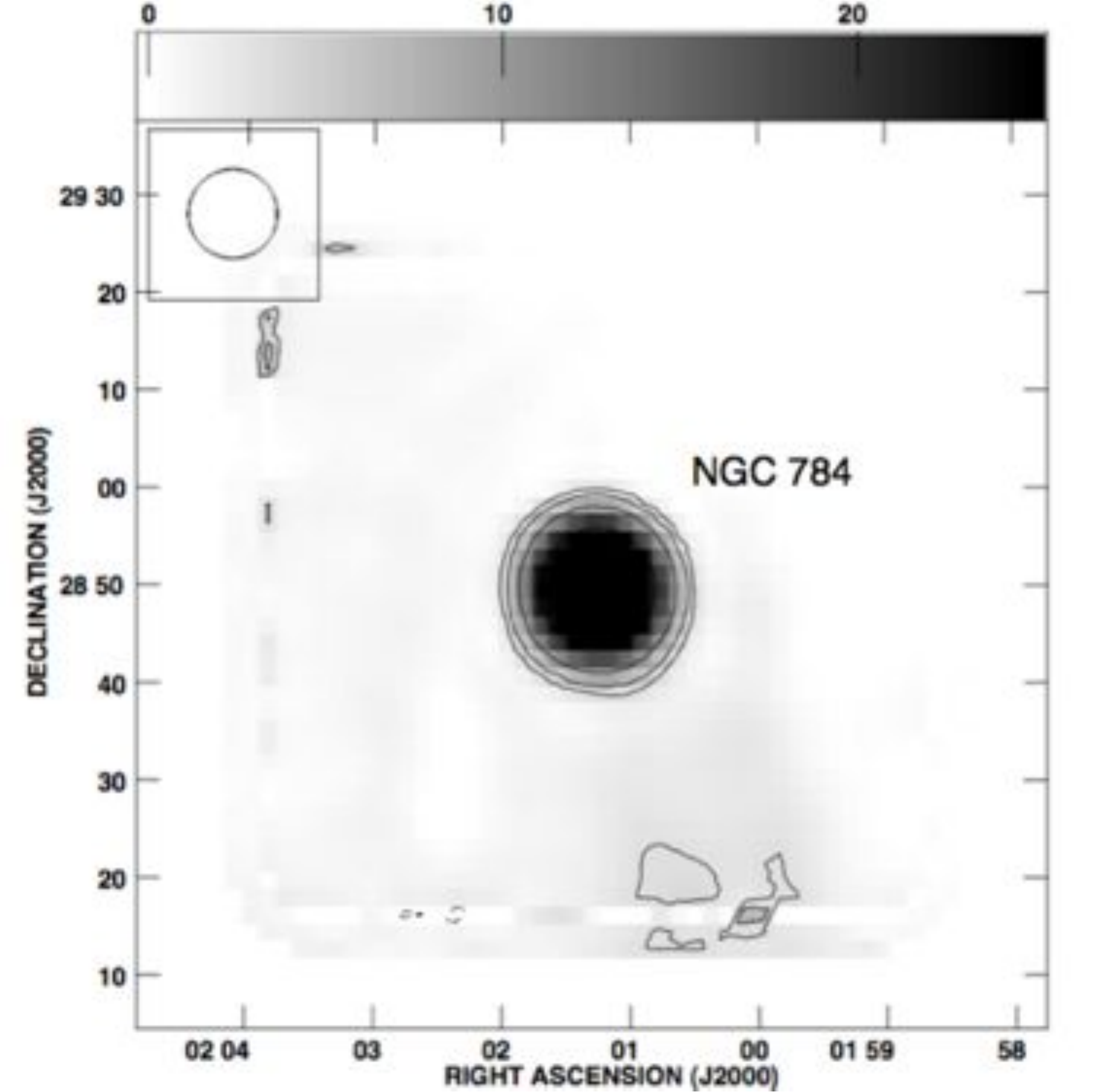} & 
\includegraphics[height=2.75in]{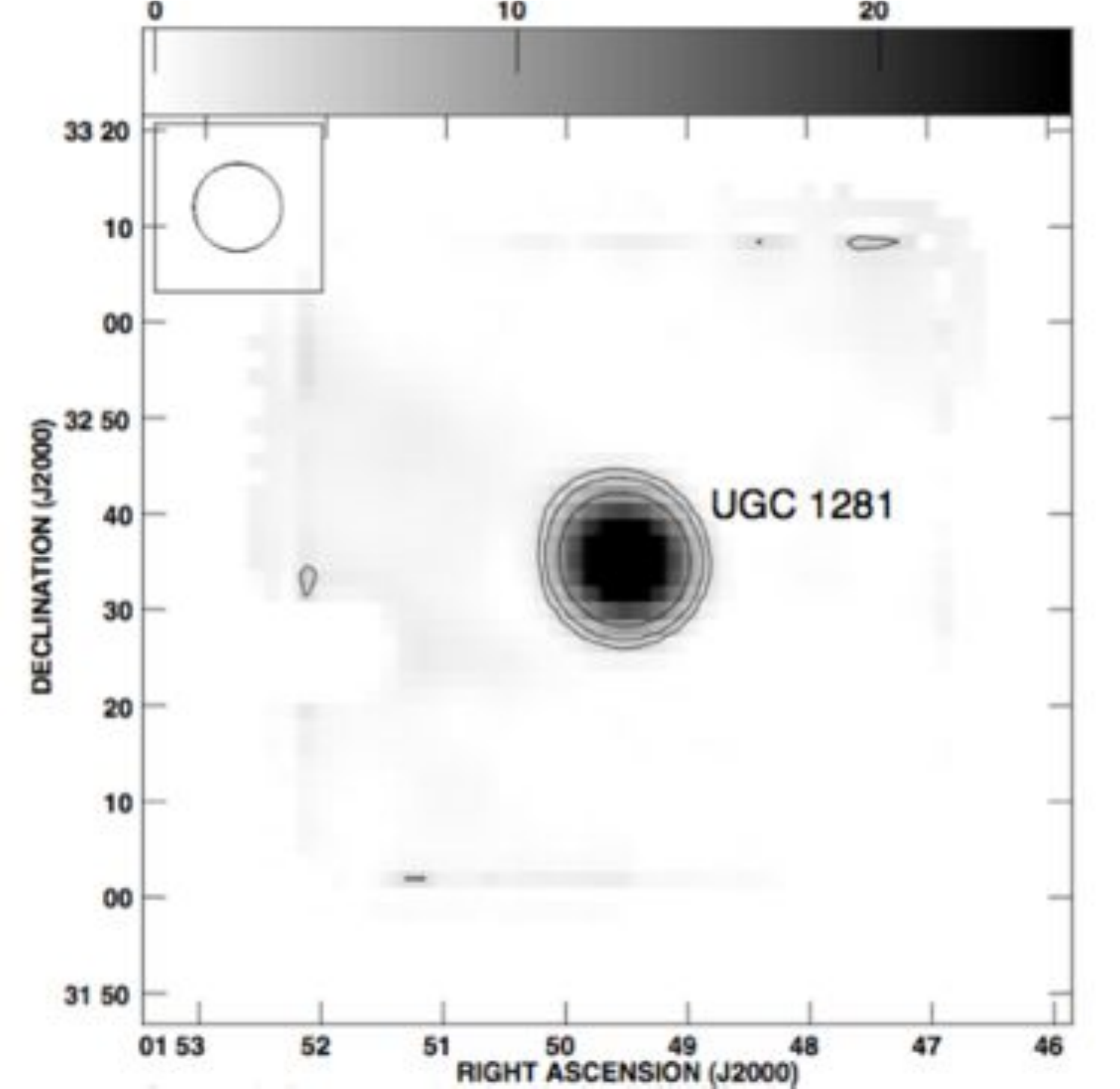} \\
\end{tabular}
\caption{\textbf{NGC 672 Group HI column density II and III.} Left: HI column density map of NGC 784. Grayscale: 0-25 Jy beam$^{-1}$ $\times$ km  s$^{-1}$. Contours: 1.0 Jy beam$^{-1}$ $\times$ km  s$^{-1}$ $\times$(-3,3,5,10,25,50,100,250. Right: HI column density map of UGC 1281. Grayscale: 0-25 Jy beam$^{-1}$ $\times$ km  s$^{-1}$. Contours: 1.0 Jy beam$^{-1}$ $\times$ km  s$^{-1}$ $\times$(-3,3,5,10,25,50,100,250). Any emission not associated with labeled galaxies is likely due to noise, since it does not have coherent velocity structure. }
\label{n784_u1281}
\end{figure}

\begin{figure}
\centering
\begin{tabular}{cc}
\includegraphics[height=2.75in]{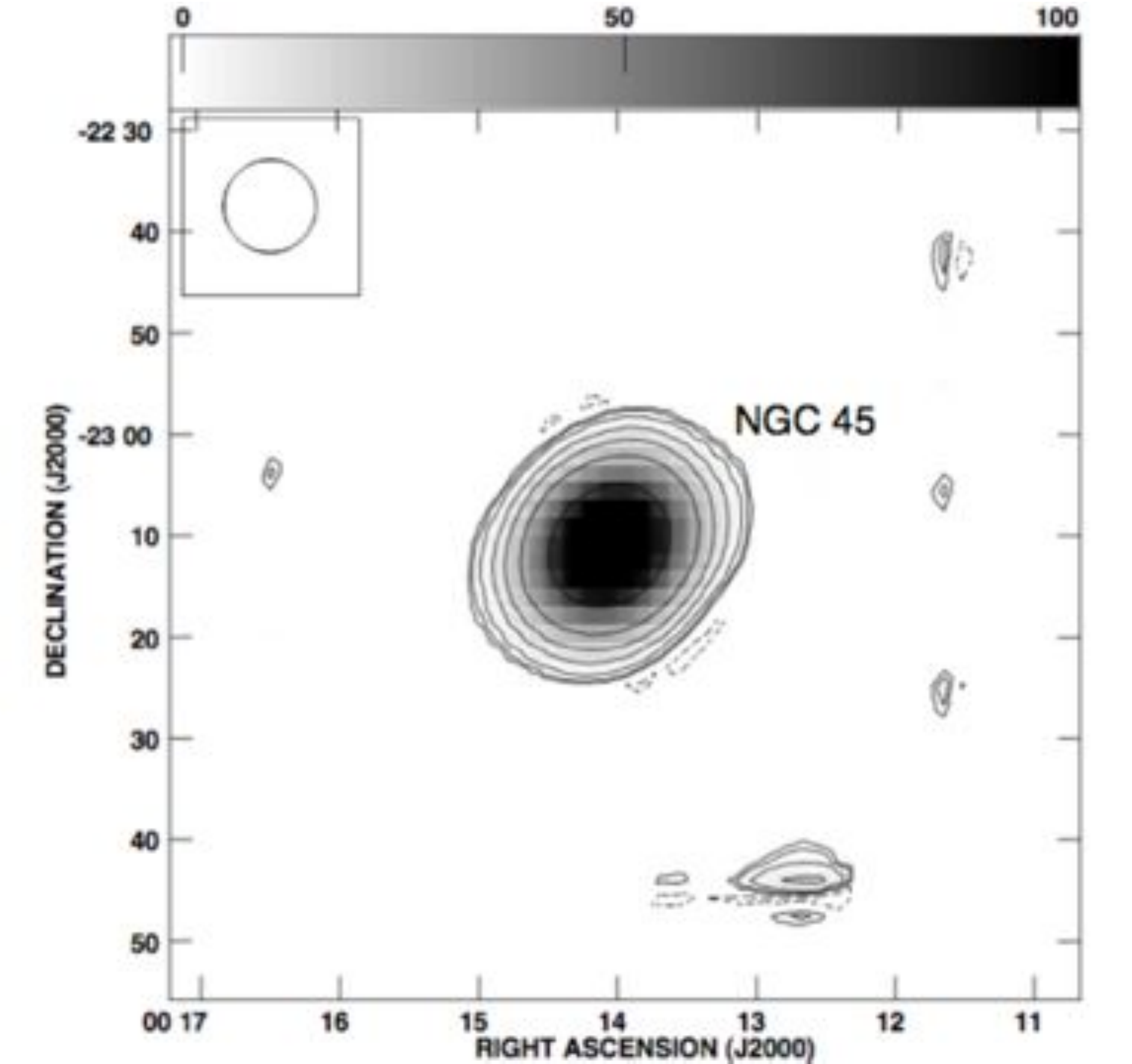} & 
\includegraphics[height=2.75in]{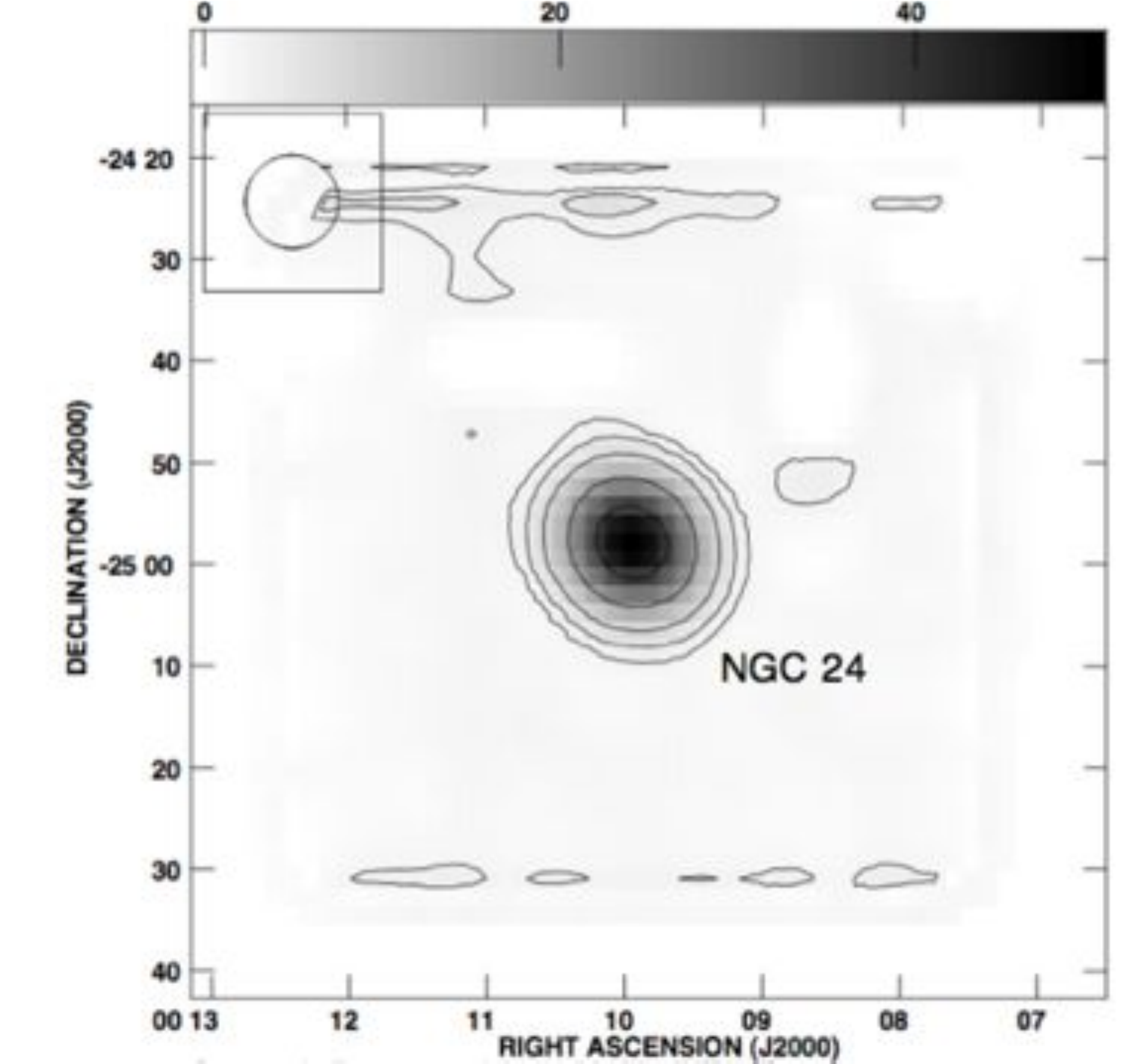} \\
\includegraphics[height=2.75in]{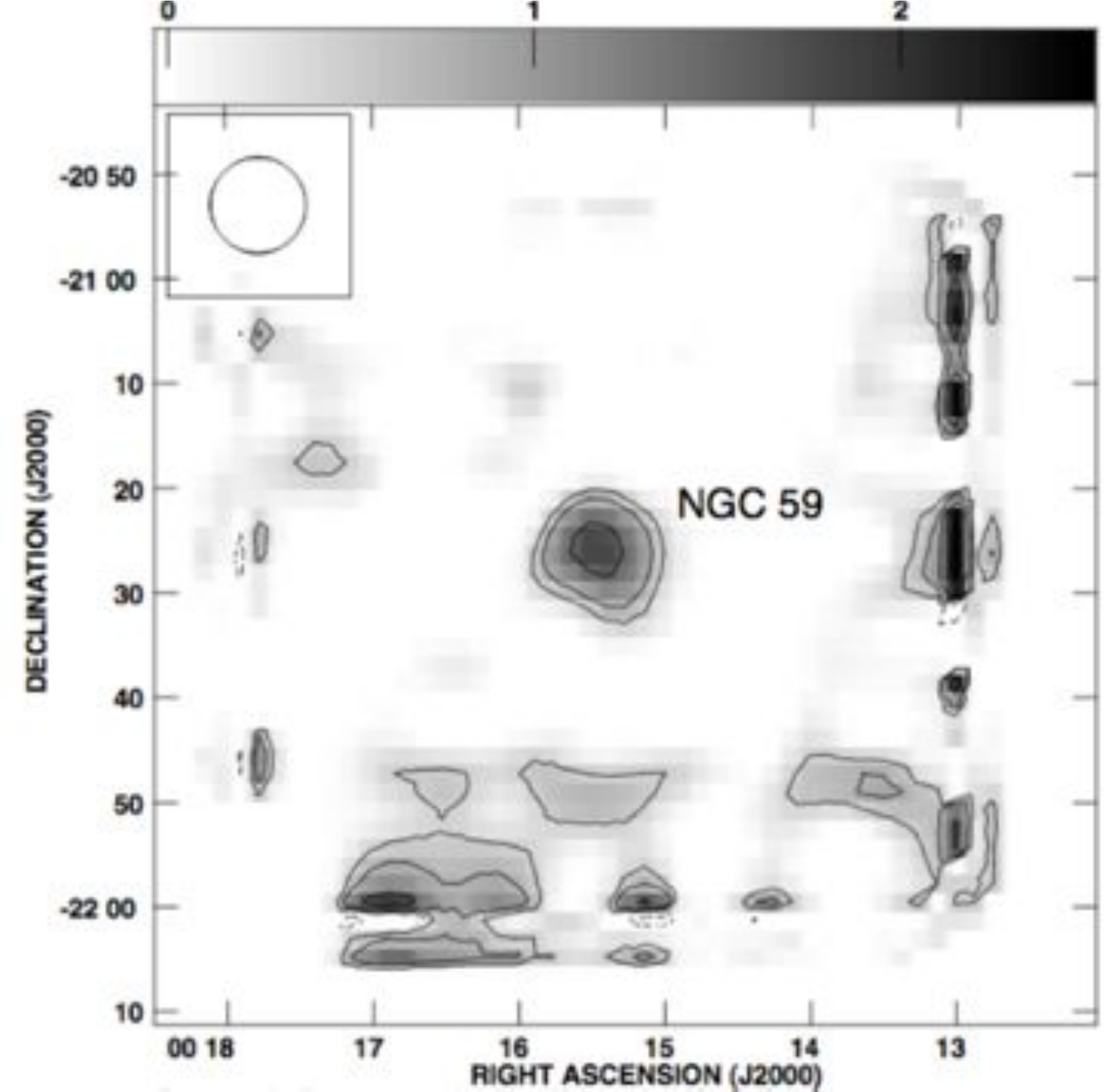} & \\
\end{tabular}
\caption{\textbf{NGC 45 group HI column density.} Any extraplanar emission is likely due to noise, since it does not have coherent velocity structure. Top Left: HI column density map of NGC 45. Grayscale: 0-100 Jy beam$^{-1}$ $\times$ km  s$^{-1}$. Contours: 0.30 Jy beam$^{-1}$ $\times$ km  s$^{-1}$ $\times$(-3,3,5,10,25,50,100,250. Top Right: HI column density map of NGC 24. Grayscale: 0-50 Jy beam$^{-1}$ $\times$ km  s$^{-1}$. Contours: 0.75 Jy beam$^{-1}$ $\times$ km  s$^{-1}$ $\times$(-3,3,5,10,25,50,100,250). Bottom Left: HI column density map of NGC 59. Grayscale: 0-2.5 Jy beam$^{-1}$ $\times$ km  s$^{-1}$. Contours: 0.15 Jy beam$^{-1}$ $\times$ km  s$^{-1}$ $\times$(-3,3,5,10,25,50,100,250)}
\label{n45_mom0}
\end{figure}

Figures \ref{mom0_canes1} and  \ref{m94_4395} show the CVn I Group observations. Figures \ref{mom0_n672} and \ref{n784_u1281} show the NGC 672 group observations. Finally, Figure \ref{n45_mom0} shows the NGC 45 group observational results. In these HI column density (moment 0) maps, we show all HI over the velocity ranges spanning each galaxy group, with galaxies labeled. Note that in some maps, there appears to be emission not associated with known galaxies. All apparent extra-galactic emission was found not to have coherent velocity structure, so this emission is likely due to noise such as undetected RFI or edge effects from the gridding function. 

\begin{figure}
\centering
\includegraphics[width=5in]{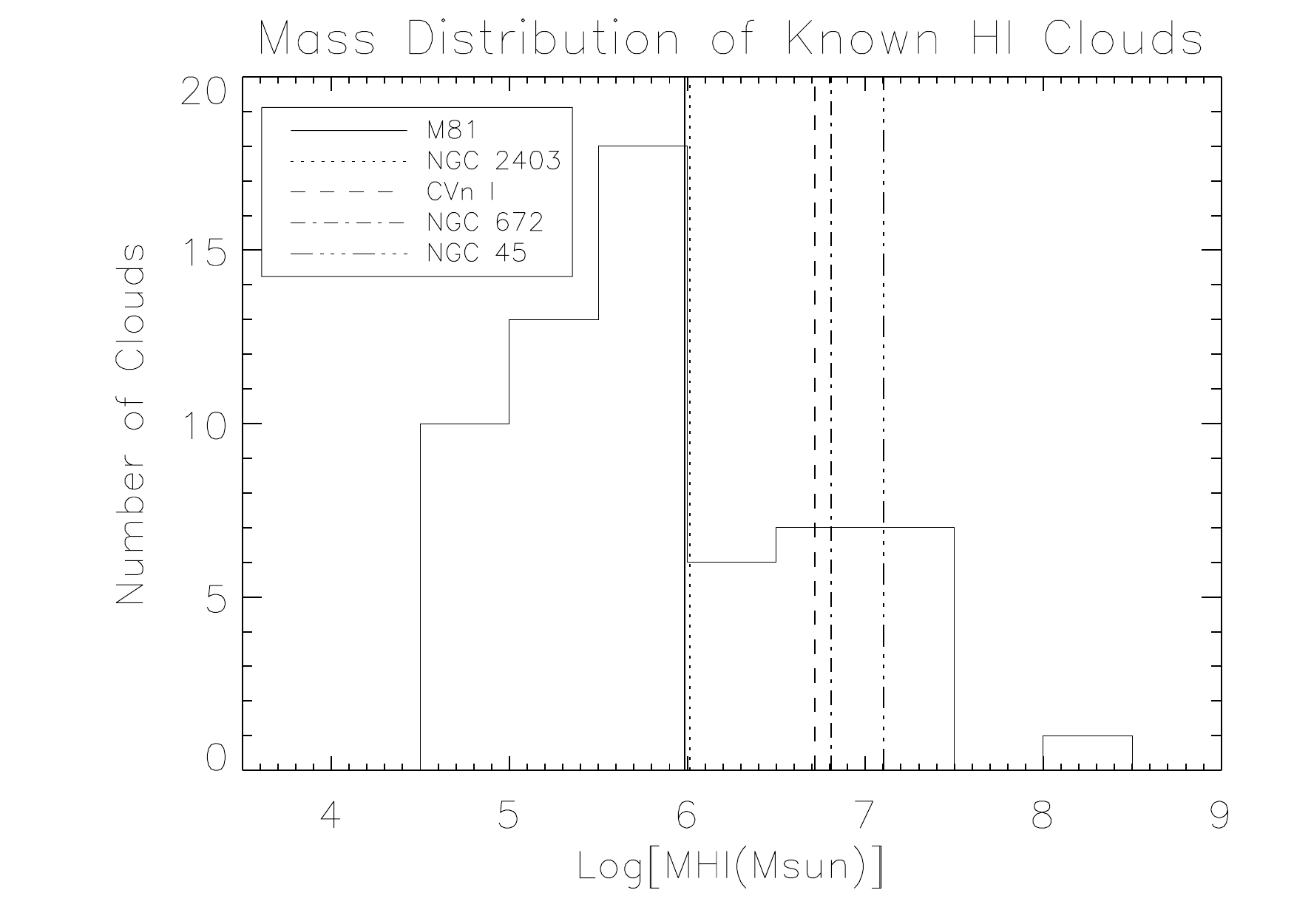}
\caption{Mass distribution of known HI clouds associated with the Milky Way and nearby galaxies M31, M33, M101, M83, and NGC 2997. Our mass detection thresholds are indicated with vertical lines. Clouds to the right of the lines would be detectable in our observations. If this population of clouds is representative of galaxy groups, our detection limit allows us to detect the most massive 11-34\% of HI clouds in the groups. Data: \citet{thilker04, grossi08,vdH88,wakker07,thom06,simon06,vanw99,put03,loc08,braun04,miller09,hess09}. } 
\label{massdist}
\end{figure}

We detected no HI clouds in the CVn I, NGC 672 or NGC 45 groups above our mass detection thresholds and within $\pm$ 700 km s$^{-1}$  and 50 kpc of group galaxies.  It is important to note that the NGC 672 and NGC 45 groups are the most distant of our observed sample, and therefore our observations are less sensitive for these groups (see Figure \ref{massdist}). More sensitive observations of all these groups would be necessary in order to better constrain their HI properties.

The observed population of HI clouds and HVCs around the Milky Way and other nearby galaxies, including those we detect in the M81 group, are likely created through a combination of galactic fountains, galaxy interactions, dark matter substructure, and cold accretion.  
The focus of this paper is to address the role of galaxy interactions and gas-embedded dark matter halos as originating mechanisms for clouds with HI masses greater than $\sim$ 10$^6$ M$_{\odot}$.  We find the role of each mechanism to be as follows.

The observed M$_{HI}$ $\ge$ 10$^6$ M$_{\odot}$  HI cloud population does not match the predictions of $\Lambda$CDM cosmological models. We have analyzed a suite of cosmological N-body simulations in order to predict the number and properties of such clouds. Our simulations predict that the number of detected HI clouds in each group should have ranged between 0-16, with an average of 3 HI clouds detected per group with our observations. The HI clouds that were detected in the M81 group are not statistically coincident in position and velocity with the simulated clouds \citep{chy11}. Unfortunately, given the mass sensitivity of our observations in the NGC 672 and NGC 45 galaxy groups, we cannot say with confidence that dark-matter associated HI clouds ought to have been detected.  It is possible that the rest of the groups in our sample are all among those major halos in the simulations that hosted no detectable HI clouds--our results would become more robust with more observed galaxy groups and/or more sensitive observations of this sample.

We find HI clouds only in the M81 galaxy group, which is currently undergoing strong interactions and ranks highest in our ``interaction evidence" ranking system. Our results therefore indicate that the majority of M$_{HI}$ $\ge$ 10$^6$ M$_{\odot}$ extragalactic HI clouds are generated through tidal stripping caused by galaxy interactions. By analogy, the Milky Way HVC population is likely generated through a combination of tidal stripping and galactic fountains (galactic fountains in the galaxy groups being un-detectable in our observations). 

Ideally, we would also address cold accretion as a possible source of HI clouds. Currently, there is no published catalog of the properties of cold accretion-generated HI clouds, and therefore we cannot make a definitive statement about the role of cold accretion. If HI clouds tracing cold accretion exist in our observed position, velocity, and mass regimes, as the simulation of \citet{keres09} predict they ought (see Table  \ref{millertable}), they should have been observed in all groups instead of only one group. If the HI clouds we have detected \textit{are} tracers of cold accretion, this result indicates that the presence or lack of cold accretion-related HI clouds is strongly environment-dependent, requiring a dense environment and/or the presence of current, strong interactions such as those in the M81 group. Tentatively, we state that the observed HI clouds are not likely to be a product of cold accretion.  Certainly, more modeling and simulations of cold accretion, especially in the very cold HI regime, are needed to properly address this question.

Future work towards determining the origin of HI clouds in galaxy groups should include high-resolution HI observations with the EVLA to determine their substructure; UV and optical observations to search for a stellar component; and comparison with simulations of the M81 Group that include all the relevant gas physics and allow for cold accretion. These studies would place stronger constraints on the origins of HI clouds in galaxy groups, and determine whether such clouds are analogs to the Milky Way HVCs. Finally, these future observations will place stronger constraints on cosmological models of galaxy formation.

\acknowledgments

KMC acknowledges the NRC Research Associateship program, the NRAO
Pre-Doctoral Fellowship program, and Vanderbilt University for funding support. 

EP acknowledges support under the Edison Memorial Graduate Training Program at the Naval Research Laboratory.

{\it Facilities:} \facility{GBT}.

\bibliographystyle{aj}

\onecolumn

\begin{table}
\begin{center}
\caption{Expected Properties of HI Clouds}
\label{millertable}
\begin{tabular}{l|c|c|c|c}
\tableline\tableline
HI Cloud Origin & Distance & Deviation Velocity & Size & Mass \\
                                   & (kpc) & (km s$^{-1}$) & (kpc) & M$_{\odot}$ \\ \tableline
Galactic Fountain\tablenotemark{a} & $<$ 10 & $<$ 150 & $<1$   & $<$ 10$^4$ \\           
Galaxy Interactions\tablenotemark{a} & 5-100 & $<$ 300 & $\sim$ 1 & 10$^5$-10$^6$ \\
Dark Matter Halos\tablenotemark{a} & 150-750 & $<$ 300 & 1-10 & 10$^6$-10$^8$ \\
Cold Accretion\tablenotemark{b}  & 30-60 & ... & ... &    10$^6$-10$^7$ \\          
\tableline
\tablenotetext{a}{From \citet{miller09}}
\tablenotetext{b}{From \citet{keres09}}
\end{tabular}
\end{center}
\end{table}

\begin{table}
\begin{center}
\caption{Galaxy Groups Within 7.5 Mpc}
\label{tenmpc}
\begin{tabular}{l|c|c|c|c}
\tableline\tableline
Name\tablenotemark{a} & $\alpha$ & $\delta$ & D & V$_{hel}$ \\
	& (J2000) & (J2000) & (Mpc) & (km s$^{-1}$) \\ \tableline
\textbf{M81}	& 10:10	& +69.1	& 3.7		& 127 \\
Maffei	& 03:49	& +68.0	& 3.1		& 171\\
Sculptor	& 00:36	& -31.0	& 2.8		& 177 \\
\textbf{NGC 2403} & 07:43	& +66.7	& 3.7 & 231 \\
\textbf{CVn I} & 12:18 & +35.8 & 4.0 & 294 \\
NGC 5128	& 13:25	& -41.6	& 4.3 	& 390 \\
\textbf{NGC 672} & 01:51	& +27.8 & 6.1	& 458 \\
\textbf{NGC 45} & 00:13 & -23.4 & 6.1 & 503 \\
\tableline
\tablenotetext{a}{Galaxy groups observed for this study are listed in bold.}
\end{tabular}
\end{center}
\end{table}

\begin{sidewaystable}
\begin{center}
\caption{M81 Group Galaxy Properties}
\label{m81tab}
\begin{tabular}{l|c|c|c|c|c|c}
\hline\hline
Name 	& $\alpha$	& $\delta$		& D        &  V$_{hel}$		& Type &  f(H$\alpha$) \\
        & (J2000)			& (J2000)			& (Mpc)    & (km s$^{-1}$) & & (log ergs cm$^{-2}$ s$^{-1}$ )	\\  \hline
M81	    & 09$^h$55$^m$33.5$^s$  & 69$^{\circ}$03$^`$06$^{``}$   & 3.65 & -34   & SA(s)ab;LINER Sy1.8  & -10.31 $\pm$ 0.02 \\
Holmberg IX 	    & 09$^h$57$^m$32.4$^s$  & 69$^{\circ}$02$^`$35$^{``}$   & 2.64 & 46    & Im 		  & -13.07 $\pm$ 0.14 \\
BK3N 	    & 09$^h$53$^m$48.5$^s$  & 68$^{\circ}$58$^`$09$^{``}$   & 4.02 & -40   & Im 		  & $<$ -15.70 \\
A0952+69    & 09$^h$57$^m$29.0$^s$  & 69$^{\circ}$16$^`$20$^{``}$   & ...  & 100   & ...		  & ... \\
KDG 61 	    & 09$^h$57$^m$02.7$^s$  & 68$^{\circ}$35$^`$30$^{``}$   & 3.45 & -135  & dE 		  & -13.41 $\pm$ 0.05 \\
M82	    & 09$^h$55$^m$53.9$^s$  & 69$^{\circ}$40$^`$57$^{``}$   & 4.55 & 203   & I0;Sbrst HII 	  & -10.09 $\pm$ 0.03 \\
NGC 3077    & 10$^h$03$^m$21.0$^s$  & 68$^{\circ}$44$^`$02$^{``}$   & 3.45 & 14    & I0 pec   HII  	  & -11.18 $\pm$ 0.04 \\
Garland	    & 10$^h$03$^m$42.7$^s$  & 68$^{\circ}$41$^`$27$^{``}$   & 3.82 & 50    &  Im  		  & ... \\
BK1N	    & 09$^h$45$^m$14.3$^s$  & 69$^{\circ}$23$^`$23$^{``}$   & 9.3  & 571   & Im 		  & ... \\
FM 1 	    & 09$^h$45$^m$10.0$^s$  & 68$^{\circ}$45$^`$54$^{``}$   & ...  & ...   & ... 		  & ... \\
BK5N 	    & 10$^h$04$^m$40.3$^s$  & 68$^{\circ}$15$^`$20$^{``}$   & 3.88 & ...   & dE 		  & ...  \\
IKN	    & 10$^h$08$^m$05.9$^s$  & 68$^{\circ}$23$^`$57$^{``}$   & ...  & ...   &  dSph       	  & ... \\
NGC 2976    & 09$^h$47$^m$15.6$^s$ & 67$^{\circ}$54$^`$49$^{``}$    & 2.83 & 3     & SAc pec HII 	  & -11.19 $\pm$ 0.06 \\
U5423	    & 10$^h$05$^m$30.6$^s$  & 70$^{\circ}$21$^`$52$^{``}$   & 7.14 & 347   & Im 		  & -12.86 $\pm$ 0.05 \\
KDG 64	    & 10$^h$07$^m$01.9$^s$  & 67$^{\circ}$49$^`$39$^{``}$   & 3.60 & -18   & Im 		  & ... \\
KK 77	    & 09$^h$50$^m$10.0$^s$  & 67$^{\circ}$30$^`$24$^{``}$   & ...  & ...   & ...	    	  & ... \\		  
IC 2574	    & 10$^h$28$^m$22.4$^s$  & 68$^{\circ}$24$^`$58$^{``}$   & 3.96 & 57    & SAB(s)m 		  & -11.23 $\pm$ 0.07 \\
\hline
\end{tabular}
\end{center}
\end{sidewaystable}

\begin{sidewaystable}
\begin{center}
\caption{NGC 2403 Group Galaxy Properties}
\label{n2403tab}
\begin{tabular}{l|c|c|c|c|c|c}
\hline\hline
Name 	& $\alpha$ 	& $\delta$ 	& D        &  V$_{hel}$ 	& Type  & f(H$\alpha$) \\
        & (J2000)			& (J2000)			& (Mpc) & (km s$^{-1}$)			        & & (log ergs cm$^{-2}$ s$^{-1}$ )	\\  \hline
NGC 2403  & 07$^h$36$^m$52.7$^s$  & 65$^{\circ}$35$^`$52$^{``}$  & 3.30   & 131    & Sc        & -10.25 $\pm$ 0.04 \\
DDO 44	  & 07$^h$34$^m$11.9$^s$  & 66$^{\circ}$52$^`$58$^{``}$  & 3.19   & ...    & dSph      & ... \\
NGC 2366  & 07$^h$28$^m$55.7$^s$  & 69$^{\circ}$12$^`$59$^{``}$  & 3.19   & 99     & IBm       & -11.01 $\pm$ 0.01 \\
UGC 4483  & 08$^h$37$^m$03.1$^s$  & 69$^{\circ}$46$^`$44$^{``}$  & 3.21   & 156    & BCD       & -12.50 $\pm$ 0.02 \\
UGC 4305  & 08$^h$19$^m$06.5$^s$  & 70$^{\circ}$43$^`$01$^{``}$  & 3.39   & 157    & Im        & -11.28 $\pm$ 0.04 \\
KDG 52    & 08$^h$23$^m$56.2$^s$  & 71$^{\circ}$01$^`$36$^{``}$  & 3.55   & 113    & Irr       & $<$ -15.30 \\
DDO 53    & 08$^h$34$^m$08.6$^s$  & 66$^{\circ}$11$^`$03$^{``}$  & 3.56   & 20     & Irr       & -12.26 $\pm$ 0.12 \\ 
VKN	  & 08$^h$40$^m$11.1$^s$  & 68$^{\circ}$26$^`$11$^{``}$  & ...    & ...    & dSph(?)   & ... \\
\hline
\end{tabular}
\end{center}
\end{sidewaystable}

\singlespace
\begin{landscape}
\begin{longtable}{l|c|c|c|c|c|c}
\caption[CVn I Group Galaxy Properties]{CVn I Group Galaxy Properties}
\label{canes1tab}
\\ \hline
Name 	& $\alpha$	& $\delta$	& D        &  V$_{hel}$ & Type  & f(H$\alpha$) \\
        & (J2000)			& (J2000)			& (Mpc)    & (km s$^{-1}$)		 & & (log ergs cm$^{-2}$ s$^{-1}$ )	\\   \hline
\endfirsthead
\multicolumn{7}{c}{{\tablename} \thetable{} -- Continued} \\
Name 	& $\alpha$ 	& $\delta$ 	& D        &  V$_{hel}$  & Type  & f(H$\alpha$) \\
        & (J2000)			& (J2000)			& (Mpc)    & (km s$^{-1}$)		 & & (log ergs cm$^{-2}$ s$^{-1}$ )	\\  \hline
 \endhead \hline \hline
 \multicolumn{7}{l}{{Continued on Next Page\ldots}} \\
\endfoot
\endlastfoot
UGC 6541 & 11:33:30 & 49:14:00  & 3.80  & 250  & Irr+Comp HII             & -12.23 $\pm$ 0.02 \\     	     
NGC 3738 & 11:35:48 & 54:31:00  &  4.24 & 229   & Irr HII	  	  & -11.81 $\pm$ 0.02 \\ 
NGC 3741 & 11:36:06 & 45:17:00  & 3.25 &  229 &	 ImIII/BCD	  	  & -12.41 $\pm$ 0.04 \\  
UGC 6817 &  11:50:48 & 38:52:00  &  3.22 & 242  & Im	  		  & -12.49 $\pm$ 0.01 \\ 
NGC 4068 & 12:04:00 & 52:35:00  & 4.53 & 210  &	IAm	  		  & -12.10 $\pm$ 0.04 \\
NGC 4163 &  12:12:06 & 36:10:00  & 2.83 & 165  & IAm BCD	  	  & -12.91 $\pm$ 0.08 \\ 
NGC 4190 & 12:13:42 & 36:38:00  & 3.14 & 228 &	Im pec	  		  & -12.33 $\pm$ 0.02 \\
UGCA 276 &  12:14:54 & 36:13:00  & 3.35 & 284  & Im:	  		  & $<$ -15.60  \\ 
NGC 4214  & 12:15:36 & 36:19:00  & 3.31 & 291  & IAB(s)m  HII   	   & -10.77 $\pm$ 0.01 \\
NGC 4244 &  12:17:30 & 37:48:00  & 4.72 & 244 &	SA(s)cd: sp HII	  	  & -11.31 $\pm$ 0.07 \\  
UGC 7321 &  12:17:30 & 22:32:00  & 3.8 & 408 &	Sd	  		  & ... \\  
IC 779	&  12:19:36 & 29:53:00  & 3.05 & 222  &	E	  		  & ... \\
IC 3308  &  12:25:18 & 26:42:00  & 4.22 & 316  & Sdm:		  	  & ... \\
NGC 4395 &  12:25:54 & 33:32:00  & 4.19 & 319  & SA(s)m:;LINER  Sy1.8	   & -11.28 $\pm$ 0.04 \\
UGCA 281 &  12:26:12 & 48:29:00  & 5.44 & 281 &	Sm pec BCDG	  	   & -11.90 $\pm$ 0.01  \\
UGC 7559 &  12:27:06 & 37:08:00  & 4.18 & 218 &	IBm	  		   & -12.45 $\pm$ 0.01  \\
UGC 7577 &  12:27:36 & 43:29:00  & 3.22 & 195 &	Im	  		   & -12.50 $\pm$ 0.05  \\
NGC 4449 &  12:28:12 & 44:05:00  & 3.38 & 207  & IBm;HII Sbrst   	   & -10.54 $\pm$ 0.02   \\
UGC 7605 &  12:28:42 & 35:42:00  & 4.18 & 310 &	Im	  		  & -12.95 $\pm$ 0.05  \\ 
UGC 7698 &  12:32:54 & 31:32:00  & 4.95 & 331  & Im	   		  & -12.23 $\pm$ 0.03  \\
UGCA 290 &  12:37:24 & 38:45:00  & 4.26 & 458  & Im pec    		  & -13.18 $\pm$ 0.04 \\  
UGCA 292 &  12:38:24 & 32:46:00  & 3.35 & 308  & ImIV-V    		  & -12.76 $\pm$ 0.01  \\
IC 3687 &  12:42:12 & 38:30:00  & 4.02 & 354  & IAB(s)m		  	  & -12.21 $\pm$ 0.03  \\ 
M94	&  12:50:54 & 41:07:00  & 5.13 & 308 & (R)SA(r)ab;Sy2 LINER	  & -10.74 $\pm$ 0.06  \\  	     
NGC 4789A & 12:54:06 & 27:09:00  & 4.15 & 374  & IB(s)m IV-V	  	   & -12.73 $\pm$ 0.05   \\  
IC 4182 &  13:05:48 & 37:36:00  & 4.59 & 321 & SA(s)m		  	   & ... \\
UGC 8215 & 13:08:00 & 46:49:00  & 4.75 & 218  &	Im	  		  & -14.39 $\pm$ 0.16	\\
NGC 5023 & 13:12:12 & 44:02:00  & 8.86 & 407  &	Scd	  		  & -12.13 $\pm$ 0.07  \\
UGC 8308 & 13:13:24 & 46:19:00  & 3.70 & 163  &	Im	  		  & -13.11 $\pm$ 0.04  \\ 
UGC 8320 & 13:14:30 & 45:55:00  & 4.31 & 192 & IBm	  		  & -12.38 $\pm$ 0.06	\\  
UGCA 342 & 13:15:06 & 42:00:00  & 6.22 & 388 &	Im	  		  & -11.87 $\pm$ 0.05	\\
NGC 5204 & 13:29:36 & 58:26:00  & 5.28 & 201 &	SA(s)m  HII 	  	  & -11.46 $\pm$ 0.03  \\ 
UGC 8508 & 13:30:48 & 54:55:00  & 2.63 & 62 &	IAm	  		  & -12.59 $\pm$ 0.06  \\  
NGC 5229 & 13:34:06 & 47:55:00  & 6.4 & 364  &	SB(s)d? sp	  	  & -12.54 $\pm$ 0.03  \\
NGC 5238 & 13:34:42 & 51:37:00  & 5.04 & 235  &	SAB(s)dm	  	  & -12.26 $\pm$ 0.06  \\
UGC 8638 & 13:39:18 & 24:47:00  & 4.11 & 274 &	Im	  		  & -12.60 $\pm$ 0.03	\\  
UGC 8651 & 13:39:54 & 40:44:00  & 3.27 & 202  &	Im	  		   & -12.61 $\pm$ 0.05  \\
UGC 8760 & 13:50:54 & 38:01:00  & 4.2 & 192 &	SAB(s)dm IV	  	   & -13.16 $\pm$ 0.11  \\ 
UGC 8833 & 13:54:48 & 35:50:00  & 3.33 & 227   & Im:	  		   & -13.36 $\pm$ 0.05  \\
UGC 9128 & 14:15:54 & 23:03:00  & 4.22 &  153 &	ImIV-V	  		  & -13.70 $\pm$ 0.16  \\
UGC 9240 & 14:24:42 & 44:31:00  & 3.13 & 150 &	IAm	  		  & -12.70 $\pm$ 0.07  \\
\hline
\end{longtable}
\end{landscape}

\begin{sidewaystable}
\begin{center}
\caption{NGC 672 Group Galaxy Properties}
\label{n672tab}
\begin{tabular}{l|c|c|c|c|c|c}
\hline\hline
Name 	& $\alpha$	& $\delta$ 	& D         &  V$_{hel}$      & Type & f(H$\alpha$) \\
           	& (J2000)		& (J2000)	          & (Mpc) & (km s$^{-1}$) &           & (log ergs cm$^{-2}$ s$^{-1}$ )	\\  \hline
NGC 672   & 01:47:54.5 & 27:25:58   & 7.35 & 422 & SB(s)cd    & -11.49 $\pm$ 0.06 \\	      
IC 1727   & 01:47:29.9 & 27:20:00   & 6.80 & 345 & SB(s)m     & -11.96 $\pm$ 0.06 \\
AGC110482 & 01:42:17.3 & 26:22:00   & 6.54 & 357 & Irr	      & ... \\
AGC111945 & 01:44:42.7 & 27:17:18   & 7.45 & 423 & Irr	      & ... \\		               
AGC111946 & 01:46:42.2 & 26:48:05   & 6.64 & 366 & Irr	      & ... \\  	        	
AGC112521 & 01:41:08.0 & 27:19:20   & 5.44 & 282 & ...	      & ... \\		               
LEDA169957 & 01:36:35.9 & 23:48:54  & 9.33 & 563 & ...	      & ... \\		        	
NGC 784   & 02:01:16.9 & 28:50:14   & 4.52 & 198 & SBdm: sp   & -11.78 $\pm$ 0.04 \\           
AGC111977 & 01:55:20.4 & 27:57:13   & 4.43 & 207 & Irr	      & ... \\		        	
AGC111164 & 02:00:10.2 & 28:49:53   & ...  & 168 & Irr 	      & ... \\ 
UGC 1281  & 01:49:32.0 & 32:35:23   & 5.03 & 156 & Sdm	      & -12.45 $\pm$ 0.07 \\
AGC122834 & 02:03:47.0 & 29:11:53   & ...  &  ...& ...	      & ... \\  	               
AGC122835 & 02:05:33.0 & 29:13:58   & ...  & ... & ...	      & ... \\		        	       
UGC 1561  & 02:04:05.1 & 24:12:30   & 9.3  & 610 & Im	      & -12.90 $\pm$ 0.04 \\    
NGC 855   & 02:14:03.6 & 27:52:36   & 9.73 & 575 & E	      & -12.23 $\pm$ 0.04 \\   
\hline
\end{tabular}
\end{center}
\end{sidewaystable}

\begin{sidewaystable}
\begin{center}
\caption{NGC 45 Group Galaxy Properties}
\label{n45tab}
\begin{tabular}{l|c|c|c|c|c|c}
\hline\hline
Name 	& $\alpha$	& $\delta$ 	& D        &  V$_{hel}$ 	& Type  & f(H$\alpha$) \\
        & (J2000)			& (J2000)			& (Mpc) & (km s$^{-1}$)			        & & (log ergs cm$^{-2}$ s$^{-1}$ )	\\  \hline
NGC 24 & 00:09:56.5 & -24:57:47 & 7.88  & 554 & SA(s)c       & -11.87 $\pm$ 0.04  \\
NGC 45 & 00:14:04.0 & -23:10:55 & 6.74  & 467 & SA(s)dm      & -11.22 $\pm$ 0.04 \\
NGC 59 & 00:15:25.1 & -21:26:40 & 5.37  & 362 & SA(rs)0-:   & -12.36 $\pm$ 0.06 \\
\hline
\end{tabular}
\end{center}
\end{sidewaystable}

\begin{sidewaystable}
\begin{center}
\caption{Interaction Index: Values}
\label{ind_ev_val}
\begin{tabular}{l|c|c|c|c|c|c}
\tableline\tableline
Galaxy Group & N$_{Tidal}$ & N$_{Tidal}$/max(N$_{Tidal}$) & N$_{AGN}$ & N$_{AGN}$/max(N$_{AGN}$) & Mean SFR & mean SFR/max(mean SFR) \\ 
		        &                      &                                                         &                       &                                                         & M$_{\odot}$ yr$^{-1}$ & \\ \tableline
\textbf{Local Group} & \textbf{2} & \textbf{0.5} & \textbf{0} & \textbf{0} & \textbf{0.47} & \textbf{1} \\
M81 & 4 & 1 & 1 & 0.5 & 0.103 & 0.22 \\
NGC 672 & 2 & 0.5 & 2 & 1 & 0.023 & 0.05 \\
CVn I & 0 & 0 & 2 & 1 & 0.027 & 0.06 \\
NGC 2403 & 0 & 0 & 0 & 0 & 0.071 & 0.15 \\
NGC 45 & 0 & 0 & 0 & 0 & 0.067 & 0.14 \\
\tableline
\end{tabular}
\end{center}
\end{sidewaystable}

\begin{table}
\begin{center}
\caption{Interaction Index: Ranked}
\label{ind_ev_rank}
\begin{tabular}{l|c|c|c|c|c}
\tableline\tableline
Galaxy Group & N$_{Tidal}$ & N$_{AGN}$ & Mean SFR & I & I Ranked \\
                         &                        &                       &                      &       &                    \\ \tableline
M81&6&5&5&16&1\\
\textbf{Local Group}&\textbf{4}&\textbf{1}&\textbf{7}&\textbf{12}&\textbf{2}\\
NGC 672 &4&6&1&11&3\\
CVn I&1&7&2&10&4\\
NGC 2403&1&1&4&6&5\\
NGC 45&1&1&3&5&6\\                        
\tableline
\end{tabular}
\end{center}
\end{table}

\begin{table}
\begin{center}
\caption{CVn I Group Observations Summary}
\label{canes1obs}
\begin{tabular}{ll}
\tableline\tableline
Area: & \\
\hskip 3mm NGC 4244 $\alpha$ range (J2000):  \dotfill & 12:05:00.0 - 12:25:00.0 \\
\hskip 3mm NGC 4244 $\delta$ range (J2000): \dotfill &  35:00:00.0 - 39:00:00.0 \\
\hskip 3mm M94 $\alpha$ range (J2000): \dotfill & 12:46:00.0 - 12:56:00.0 \\
\hskip 3mm M94 $\delta$ range (J2000): \dotfill &   40:00:00.0 - 42:00:00.0 \\
\hskip 3mm NGC 4395 $\alpha$ range (J2000): \dotfill & 12:21:00.0 - 12:31:00.0 \\
\hskip 3mm NGC 4395 $\delta$ range (J2000): \dotfill &  32:30:00.0 - 34:30:00.0 \\
Observations: & \\
\hskip 3mm Center Frequency (MHz): \dotfill & 1420.4 \\
\hskip 3mm Bandwidth (MHz): \dotfill & 12.5 \\
\hskip 3mm Velocity Range (Heliocentric, km s$^{-1}$)): \dotfill & -1315 - 1320 \\
\hskip 3mm Channel Width (kHz): \dotfill & 24.4 \\
\hskip 3mm Velocity Resolution (km s$^{-1)}$ \dotfill & 5.2 \\
Integration time (hours): & \\
\hskip 3mm NGC 4244: \dotfill & 42.2 \\
\hskip 3mm M94: \dotfill & 9.8 \\
\hskip 3mm NGC 4395: \dotfill & 5.7 \\
Typical RMS noise (mK/channel): & \\
\hskip 3mm NGC 4244: \dotfill & 26.3 \\
\hskip 3mm M94: \dotfill & 21.2 \\
\hskip 3mm NGC 4395: \dotfill & 32.6 \\
Sensitivity to HI (1$\sigma$): & \\
M$_{HI}$ ($\times$ 10$^5$ M$_{\odot}$) & \\
\hskip 3mm NGC 4244: \dotfill & 2.40 \\
\hskip 3mm M94: \dotfill & 4.16  \\
\hskip 3mm NGC 4395: \dotfill & 4.27 \\
Beam size (kpc): & \\
\hskip 3mm NGC 4244: \dotfill & 12.7 \\
\hskip 3mm M94: \dotfill & 13.9  \\
\hskip 3mm NGC 4395: \dotfill & 11.3 \\
\tableline
\end{tabular}
\end{center}
\end{table}

\begin{table}
\begin{center}
\caption{NGC 672 Group Observations Summary}
\begin{tabular}{ll}
\tableline\tableline
Area: & \\
\hskip 3mm NGC 672 $\alpha$ range (J2000):  \dotfill & 01:38:00.0 - 01:52:00.0 \\
\hskip 3mm NGC 672 $\delta$ range (J2000): \dotfill &  28:00:00.0 - 26:00:00.0 \\
\hskip 3mm NGC 784 $\alpha$ range (J2000): \dotfill & 01:58:00.0 - 02:04:00.0 \\
\hskip 3mm NGC 784 $\delta$ range (J2000): \dotfill &   29:30:00.0 - 28:10:00.0 \\
\hskip 3mm UGC 1281 $\alpha$ range (J2000): \dotfill &  01:47:00.0 - 01:15:00.0 \\
\hskip 3mm UGC 1281 $\delta$ range (J2000): \dotfill &  33:10:00.0 - 32:00:00.0 \\
Observations: & \\
\hskip 3mm Center Frequency (MHz): \dotfill & 1418.4 \\
\hskip 3mm Bandwidth (MHz): \dotfill & 12.5 \\
\hskip 3mm Velocity Range (Heliocentric, km s$^{-1}$)): \dotfill &   -890 - 1740 \\
\hskip 3mm Channel Width (kHz): \dotfill & 24.4 \\
\hskip 3mm Velocity Resolution (km s$^{-1)}$ \dotfill & 5.2 \\
Integration time (hours): & \\
\hskip 3mm NGC 672: \dotfill & 36.4 \\
\hskip 3mm NGC 784: \dotfill & 5.3 \\
\hskip 3mm UGC 1281: \dotfill & 5.7 \\
Typical RMS noise (mK/channel): & \\
\hskip 3mm NGC 672: \dotfill & 5.92 \\
\hskip 3mm NGC 784: \dotfill & 12.3 \\
\hskip 3mm UGC 1281: \dotfill & 12.8 \\
Sensitivity to HI (1$\sigma$): & \\
M$_{HI}$ ($\times$ 10$^5$ M$_{\odot}$) & \\
\hskip 3mm NGC 672: \dotfill & 1.99 \\
\hskip 3mm NGC 784: \dotfill & 1.88  \\
\hskip 3mm UGC 1281: \dotfill & 2.41 \\
Beam size (kpc): & \\
\hskip 3mm NGC 672: \dotfill & 19.8 \\
\hskip 3mm NGC 784: \dotfill & 12.2  \\
\hskip 3mm UGC 1281: \dotfill & 13.6 \\
\tableline
\end{tabular}
\end{center}
\end{table}

\begin{table}
\begin{center}
\caption{NGC 45 Group Observations Summary}
\label{n45obs}
\begin{tabular}{ll}
\tableline\tableline
Area: & \\
\hskip 3mm NGC 45 $\alpha$ range (J2000):          \dotfill & 00:11:00.0 - 00:17:00.0 \\
\hskip 3mm NGC 45 $\delta$ range (J2000): \dotfill & -22:30:00.0 - -23:50:00.0 \\
\hskip 3mm NGC 24 $\alpha$ range (J2000): \dotfill & 00:07:00.0 - 00:13:00.0 \\
\hskip 3mm NGC 24 $\delta$ range (J2000): \dotfill & -24:20:00.0 - -25:40:00.0 \\
\hskip 3mm NGC 59 $\alpha$ range (J2000): \dotfill & 00:13:00.0 - 00:18:00.0 \\
\hskip 3mm NGC 59 $\delta$ range (J2000): \dotfill &  -20:50:00.0 - -22:10:00.0 \\
Observations: & \\
\hskip 3mm Center Frequency (MHz): \dotfill & 1420.4 \\
\hskip 3mm Bandwidth (MHz): \dotfill & 12.5 \\
\hskip 3mm Velocity Range (Heliocentric, km s$^{-1}$)): \dotfill &  -1315 - 1320 \\
\hskip 3mm Channel Width (kHz): \dotfill & 24.4 \\
\hskip 3mm Velocity Resolution (km s$^{-1)}$ \dotfill & 5.2 \\
Integration time (hours): & \\
\hskip 3mm NGC 45: \dotfill & 11.3 \\
\hskip 3mm NGC 24: \dotfill & 12.9 \\
\hskip 3mm NGC 59: \dotfill & 16.5 \\
Typical RMS noise (mK/channel): & \\
\hskip 3mm NGC 45: \dotfill & 13.9 \\
\hskip 3mm NGC 24: \dotfill & 10.2 \\
\hskip 3mm NGC 59: \dotfill & 10.9 \\
Sensitivity to HI (1$\sigma$): & \\
M$_{HI}$ ($\times$ 10$^5$ M$_{\odot}$) & \\
\hskip 3mm NGC 45: \dotfill & 4.72 \\
\hskip 3mm NGC 24: \dotfill & 4.72  \\
\hskip 3mm NGC 59: \dotfill & 2.19 \\
Beam size (kpc): & \\
\hskip 3mm NGC 45: \dotfill & 21.3 \\
\hskip 3mm NGC 24: \dotfill & 18.2  \\
\hskip 3mm NGC 59: \dotfill & 14.5 \\
\tableline
\end{tabular}
\end{center}
\end{table}

\begin{table}
\begin{center}
\caption{Mass Detection Thresholds}
\label{masscomptable}
\begin{tabular}{l|c|c}
\hline\hline
Galaxy Group	& Average Detection Threshold & Fraction of Detectable HI Clouds  \\
                            & (5$\sigma$, $\times$ 10$^6$ M$_{\odot}$)                  &                \\ \hline
M81			& 0.97 & 0.34 \\
NGC 2403	& 1.03  & 0.32 \\
CVn I		& 5.23  & 0.23 \\
NGC 672		& 6.43  & 0.21 \\
NGC 45		& 12.7  & 0.11 \\
\hline
\end{tabular}
\end{center}
\end{table}

\begin{table}
\begin{center}
\caption{Numerical Simulation Predictions\tablenotemark{a}}
\label{simtable}
\begin{tabular}{l|c|c|c|c|c}
\hline\hline
Galaxy Group	& Probability of & $<N_{clouds}>$ &  $<M_{HI}>$ &  $<V_{cloud}-V_{galaxy}>$ & $<R_{cloud}-R_{galaxy}>$ \\
                            &  detectable HI clouds                  &  &  ($\times$ 10$^6$ M$_{\odot}$)  & (km s$^{-1}$) & (kpc)            \\ \hline
M81			& 0.97 & 6 & 7.4  & 172 & 37 \\
NGC 2403	& 0.97      & 5  & 7.0 & 209 & 36 \\
CVn I		& 0.67 & 2  & 19.8 & 201 & 35 \\
NGC 672		& 0.50 &  2 & 11.4 & 251 & 35 \\
NGC 45		& 0.38 &  1 & 39.3 & 242 & 33 \\
\hline
\end{tabular}
\tablenotetext{a}{All values calculated within $\pm$ 50 kpc  and $\pm$ 700 km s$^{-1}$ of central halo, and averaged over all numerical simulations. Properties of HI clouds are calculated for the major dark-matter halos that host detectable HI clouds.}
\end{center}
\end{table}

\begin{table}
\begin{center}
\caption{M81 Filament HI Cloud Properties}
\label{fil_cloudprop}
\begin{tabular}{l|c|c|c|c|c|c}
\hline\hline
Cloud \tablenotemark {a}	& Coordinate Designation  & $\Delta$D$_{M81}$ & T$_{peak}$	& $\Delta$V$_{M81}$	&$\sigma_v$  & M$_{HI}$ ($\frac{D}{3.63 Mpc})^{-2}$ \\ 
		&                       & (kpc) &(K)	      & (km s$^{-1}$)		& (km s$^{-1}$) & $\times$ 10$^{6}$M$_{\odot}$\\ \hline
C08-1 & GBC J095007.2+695556  & 65 & 0.11 & 202 & 50 & 14.7 \\
C08-2 & GBC J100250.7+681949  & 63 & 0.10 & 68 & 55 & 22.5 \\
C08-3 & GBC J095244.9+681250  & 57 & 0.12 & 48 & 82 & 26.7 \\
C08-4 & GBC J100145.0+691631  & 38 & 0.30 & 108 & 28 & 83.7 \\
C08-5 & GBC J095506.3+692205 & 21 & 0.07 & 317 & 36 & 6.9 \\
C10-1   & GBC J092635.8+702850 & 186 & 0.13 & 144  & 57 &3.6  \\
C10-2   & GBC J101926.3+675222 & 161 & 0.20 & 72 & 26 & 3.4\\
C10-3   & GBC J091952.7+680937 & 218 & 0.17 & 74 & 42 & 5.1\\
C10-4  & GBC J102239.1+684057 & 159 & 0.30 & 103 & 68 & 12.0 \\ \hline \hline
\textbf{Mean Values} & ... & 108 & 0.17 & 126 & 49 & 19.8 \\
\textbf{Simulated Mean Values}\tablenotemark{b} & ... & 170 & ... & 250 & ... & 10.8  \\
\hline
\tablenotetext{a}{C08 indicates clouds from \citet{chy08}, C09 indicates clouds from \citet{chy09}, and C10 indicates clouds from \citet{chy11}.}
\tablenotetext{b}{From \citet{chy11}: calculated in a region matching the large size ($\pm$ 250 kpc) and velocity range ($\pm$ 900 km s$^{-1}$) of the M81 Filament observation.}
\end{tabular}
\end{center}
\end{table}

\end{document}